\documentclass[%
 reprint,
 amsmath,amssymb,superscriptaddress,floatfix,
pra,
]{revtex4-2}
\usepackage{CJK}
\usepackage{amsmath}
\usepackage{mathtools}
\usepackage{graphicx}  
\usepackage{physics}
\usepackage{graphicx}
\usepackage{dcolumn}
\usepackage{bm}

\usepackage{amssymb}
\usepackage{epstopdf}
\usepackage{inputenc}

\begin{document}

\title{A thermodynamic description of the near- and far-field intensity patterns emerging from multimode nonlinear waveguide arrays}

\author{Mahmoud A. Selim}
\author{Fan O. Wu}%
\affiliation{CREOL, College of Optics and Photonics, University of Central Florida Orlando, Florida 32816-2700, USA}
\author{Huizhong Ren}%
\affiliation{CREOL, College of Optics and Photonics, University of Central Florida Orlando, Florida 32816-2700, USA}

\author{Mercedeh Khajavikhan}%
\affiliation{Department of Electrical and Computer Engineering, University of Southern California, Los Angeles, California 90007, USA}

\author{Demetrios Christodoulides}%
\email{demetri@creol.ucf.edu}
\affiliation{CREOL, College of Optics and Photonics, University of Central Florida Orlando, Florida 32816-2700, USA}

\begin{abstract}
Nonlinear highly multimode photonic systems are ubiquitous in optics. Yet, the sheer complexity arising from the action of nonlinearity in multimode environments has posed theoretical challenges in describing these systems. In this work, we deploy concepts from optical thermodynamics to investigate the near- and far-field emission intensity patterns emerging from nonlinear waveguide arrays. An exact equation dictating the response of a nonlinear array is derived in terms of the system’s invariants that act as extensive thermodynamic variables. In this respect, the near- and far-field characteristics emerging from a weakly nonlinear waveguide lattice are analytically addressed. We show that statistically, these patterns and the resulting far-field brightness are governed by the optical temperature and its corresponding chemical potential. The extensivity associated with the entropy of such configurations is discussed. The thermodynamic results presented here were found to be in good agreement with numerical simulations obtained from nonlinear coupled-mode theory.

\end{abstract}

\maketitle

\section{\label{sec:level1}INTRODUCTION\protect\\ }
During the last few years, nonlinear multimode optical systems have attracted considerable attention  \cite{renninger2013optical,xiong2017kuznetsov,bao2019laser,xia2021nonlinear,zhang2021solitary}. Along these lines, a host of new effects have been successfully observed. These include geometric parametric instabilities \cite{krupa2016observation,longhi2003modulational}, beam self-cleaning  \cite{krupa2017spatial}, spatiotemporal mode-locking \cite{wright2017spatiotemporal}, and supercontinuum generation \cite{lopez2016visible}, to mention a few. The multitude of modes supported by these systems provides not only new physical settings but also rich and complex environments where new classes of nonlinear spatiotemporal interactions can emerge \cite{wright2015controllable}. To some extent, this convoluted and chaotic nonlinear energy exchange among optical modes is akin to that encountered in many-body problems \cite{ramos2020optical,shi2021controlling}, the investigation of which typically requires considerable computational power. Quite recently, a thermodynamic formalism has been developed that can describe in an effortless manner the classical behavior of highly multimode weakly nonlinear bosonic systems whose dynamics involve two conserved quantities \cite{parto2019thermodynamic,wu2019thermodynamic}. This approach is universal and applicable to both multimode waveguide and cavity arrangements, irrespective of the type of nonlinearity used, as long as ergodicity is at play \cite{makris2020statistical}.

An important and versatile family of multimode structures is that pertaining to one- and two- dimensional photonic waveguide lattices. In the tight-binding regime, these systems are known to display wave dynamics that are mathematically isomorphic to electron transport in solid-state physics \cite{yablonovitch1987inhibited}. Some of these effects include, for instance, Bloch oscillations \cite{pertsch1999optical,peschel1998optical}, topologically protected edge states and supersymmetric dynamics \cite{rechtsman2013photonic}, highly degenerate flat band Lieb lattices \cite{xia2018unconventional}, Zenner tunneling \cite{trompeter2006visual}, and dynamic localization processes \cite{szameit2009polychromatic}. Interestingly, when such a lattice is nonlinear, the propagation dynamics can be fundamentally altered, leading to discrete soliton formation and the appearance of novel topological phases  \cite{xia2021nonlinear,lederer2008discrete}. On the other hand, in the weakly nonlinear regime, as the number of modes increases, these self-organized patterns tend to dissolve because of nonlinear multi-wave mixing effects \cite{wu2019thermodynamic,onorato2015route}-an inherently complex process that has so far remained unexplored. In general, the near- and far-field intensity patterns at the output are known to play an important role in several applications \cite{dostart2020serpentine,partovi1999high}. As such, of interest will be to develop a formalism that could capture the macroscopic behavior of these weakly nonlinear waveguide arrays using notions from statistical mechanics. Along these lines, previous studies have shown that an optical Sackur-Tetrode equation can be utilized to describe nonlinear chain networks under thermal equilibrium conditions-an equation that explicitly provides in closed form the optical temperature and chemical potential of such systems \cite{wu2020entropic,rumpf2007growth}. In general, this formalism is based on the assumption that these networks are highly multimoded-an aspect that is not always met in deploying such a statistical approach. In addition, to our knowledge, no analytical expression has been obtained that could describe the nonlinear waveguide array’s near- and far-field intensity patterns at thermal equilibrium. It is important to emphasize that the optical thermodynamic theory outlined here deals with photon-photon interactions mediated by Kerr nonlinearities. This is in contrast to  photon-matter interactions \cite{klaers2010bose,deng2002condensation,kasprzak2006bose,jamadi2016polariton,sun2017bose,walker2018driven},  where the temperature is dictated by the temperature of the thermal bath or environment. As we will see, the optical temperature in the multimode nonlinear arrangements discussed here is uniquely determined by the initial optical power, ``internal energy'', and the number of modes.

In this manuscript, by means of optical thermodynamics, we provide a methodology for analytically predicting and describing the near- and far-field radiation patterns emerging from waveguide chain networks. We formally derive the main equations for the aforementioned nonlinear waveguide systems in terms of the extensive variables involved (total power, internal energy, and the number of modes) \cite{wu2019thermodynamic} and the intensive quantities associated with the optical temperature and chemical potential. In addition, we clarify the mode range over which the system is entropically extensive. In general, we find that the temperature and chemical potential govern the near- and far-field intensity distributions via a Rayleigh-Jeans law. In all cases, the analytical results from this optical thermodynamic theory have been augmented with numerical simulations. 

\section{GENERAL ANALYSIS}
We begin our analysis by considering an array of $M$ coupled waveguides, as shown in Fig. 1. The system is governed by the following normalized discrete nonlinear Schr\"{o}dinger  equation:
\begin{eqnarray}
\frac{i d{\Psi }_{n}}{dz}+\kappa \left({\Psi }_{n+1}+{\Psi }_{n-1}\right)+{\left|{\Psi }_n\right|}^2{\Psi }_n=0,
\label{eq:1}
\end{eqnarray}
\begin{figure}
\includegraphics*[width=3.20in, height=2in]{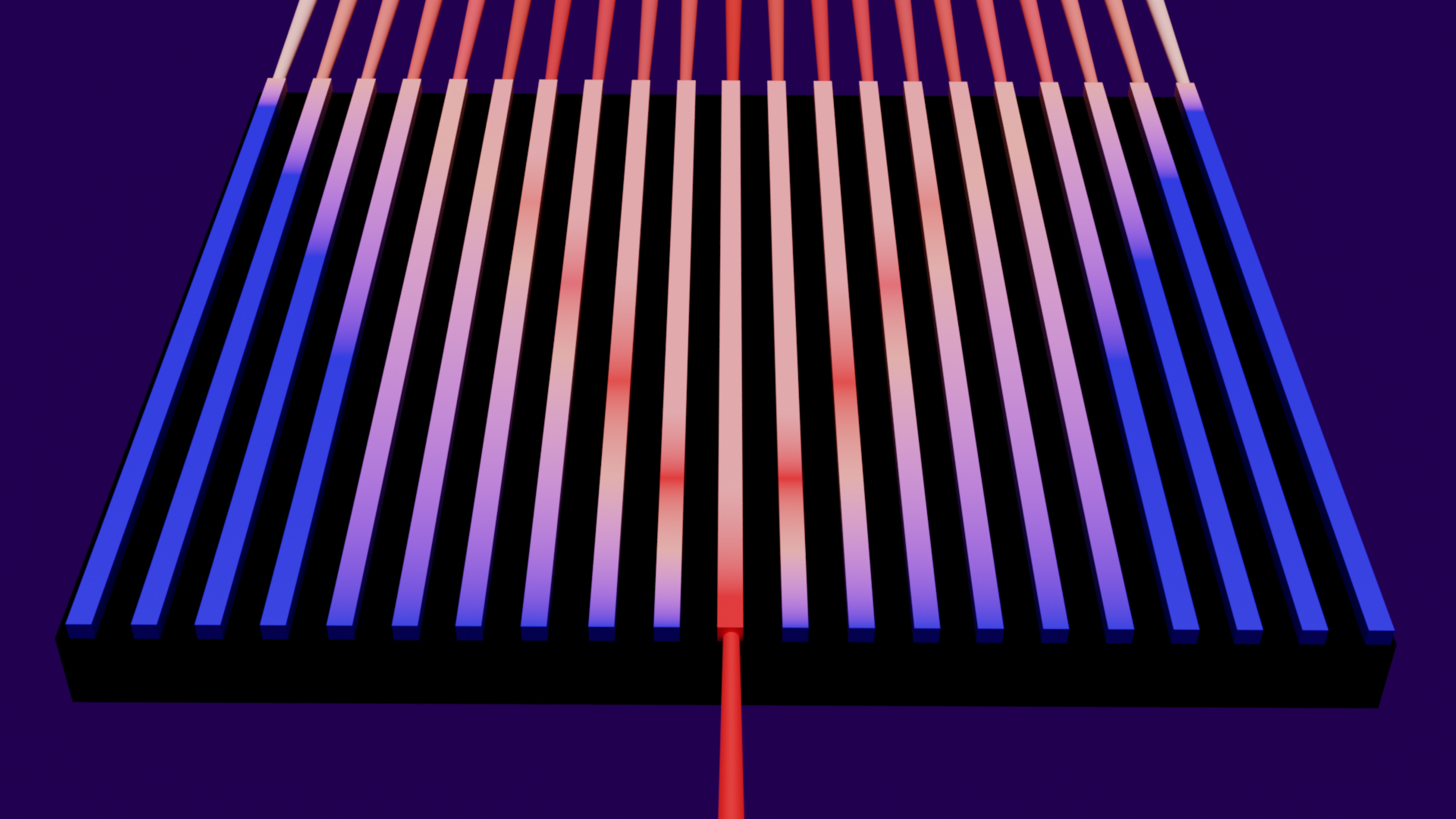}
\caption{\label{fig:1}An on-chip photonic lattice consisting of several identical waveguide channels. Each channel is designed to be single-moded at the operating wavelength.}
\end{figure}
where \(\ {\Psi }_{n}\) is the complex amplitude of the optical field at site \(n\), and \(\kappa\) is the normalized coupling coefficient. In the linear regime, this arrangement supports \(M\) eigenmodes which can be obtained from the following Hamiltonian problem \(\ H _{L}\ket{\psi_{j}}={\varepsilon}_{j}\ket{{\psi}_{j}}\), where \(H_L\) is the linear Hamiltonian of the system, \(\varepsilon_{j}\) represents the eigenvalue associated with the eigenmode \(\ket{{\psi}_{j}}\). For convenience, we assume \({\varepsilon}_{1}\leq{\varepsilon}_{2}\leq{\varepsilon}_{3}\leq \ldots \leq{\varepsilon}_{M}\), where \({\varepsilon}_{1}\) and \({\varepsilon}_{M}\) denote the highest-order and lowest-order mode, respectively. In this case, the optical field can be expressed as a superposition of these eigenstates, i.e., \(\ket{{\Psi}_{}}=\sum_{j=1}^{M} c_{j}{\psi }_{j} \).  Under weakly nonlinear conditions, both the optical power \(\mathcal{P}=\sum_{j=1}^{M} |c_{j}|^{2} \) and the internal energy \(U=-\sum^M_{j=1}{{{\varepsilon }_j\left|c_j\right|}^2}\) are conserved and are determined from the initial excitation conditions \({\left|c_{j0}\right|}^2={\left|\left\langle {\psi }_j\mathrel{\left|\vphantom{{\psi }_i {{\Psi}}_0}\right.\kern-\nulldelimiterspace}{{\Psi(0)}}\right\rangle \right|}^2\). In this respect, the sole role of nonlinearity is to chaotically reshuffle via multi-wave mixing processes \cite{boyd2008nonlinear} the optical power among modes ${\left|c_j\right|}^2,$ while respecting the constraints of power and energy conservation manifolds. Such ergodic dynamics lead to a thermal equilibrium state that obeys a Rayleigh-Jeans distribution ${\left|c_j\right|}^2=-T/{(\varepsilon }_j+\mu )$ \cite{wu2019thermodynamic,zakharov1985hamiltonian,picozzi2014optical}, where $T$ and $\mu $ represent the optical temperature and chemical potential, respectively. Interestingly, these thermodynamic properties are linked to each other via a global equation of state $U-\mu \mathcal{P}=MT$ \cite{wu2019thermodynamic}, and thus the final temperature and chemical potential can be uniquely determined from initial conditions \cite{parto2019thermodynamic}.

The column elements of the supermodes $\left.|{\psi }_j\right\rangle \ $associated  with this one-dimensional waveguide array can be expressed as ${\psi }_j(n)=\sqrt{2/(M+1)}{\mathrm{sin} \left(\frac{nj\pi }{M+1}\right)\ }$ \cite{butler1984coupled,kapon1984supermode}. The corresponding eigenvalue of each supermode is given by ${\varepsilon }_j=2\kappa {\mathrm{cos} \left(\frac{j\pi }{M+1}\right)\ }$ \cite{kapon1984supermode}. In this case, upon thermal equilibrium, the total optical power in the system can be rewritten as
\begin{eqnarray}
\mathcal{P}=-\sum^M_{j=1}{\frac{T/\mu }{A\mathrm{cos}\left(\frac{j\pi }{M+1}\right)+1}},
\label{eq:2}
\end{eqnarray}

\begin{figure}
\includegraphics*[width=3in, height=4in]{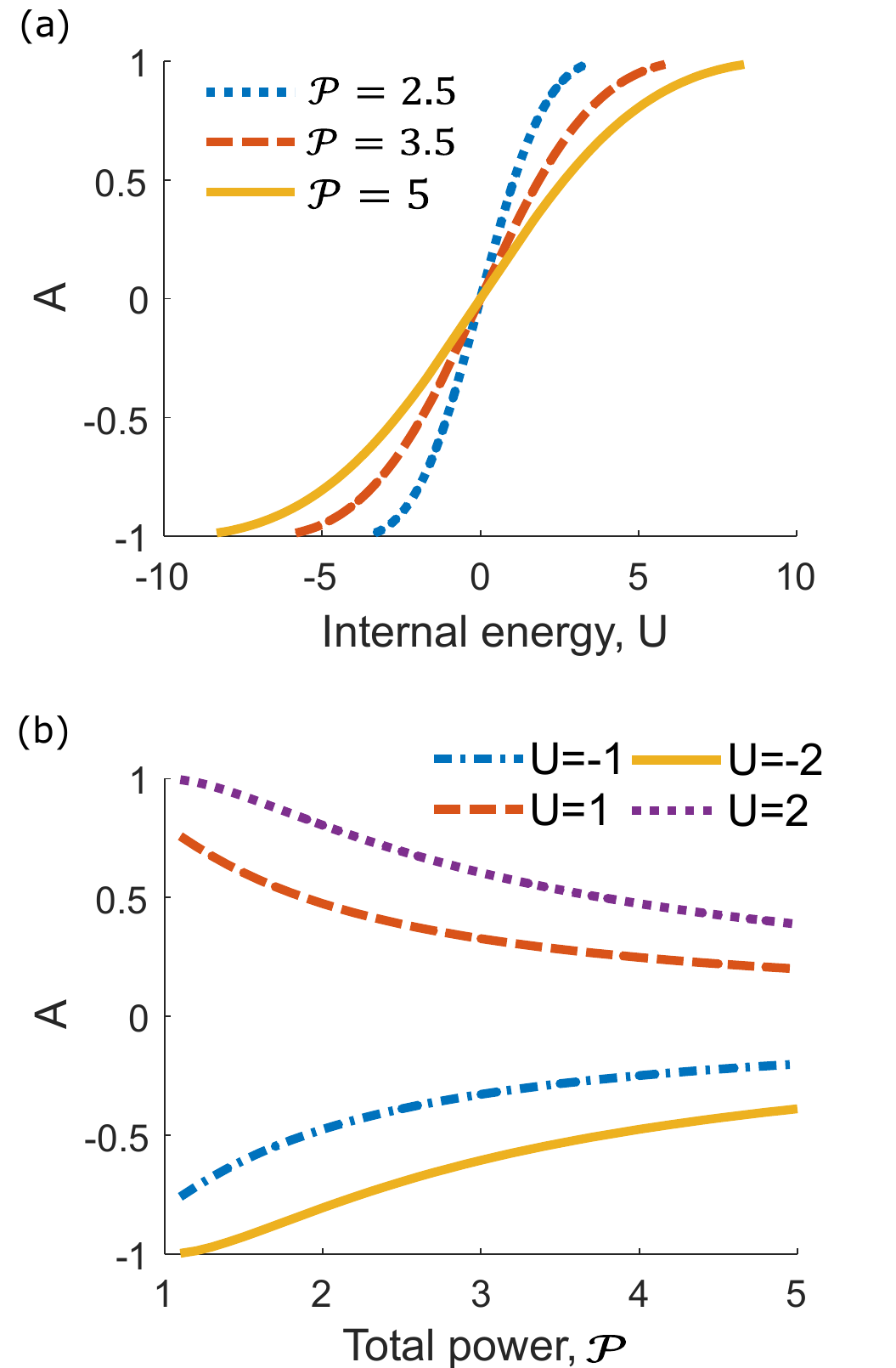}
\caption{\label{fig:2}(a) Parametric curves depicting the dependence of the quantity ${A}$ on the internal energy $U$ for various optical power levels $\mathcal{P}$. (b) Dependence of ${A}$ on $\mathcal{P}$ for different values of $U$.}
\end{figure}

where $A=\frac{2\kappa }{\mu }$. The quantity\textit{ }$|A|$\textit{ }is always less than unity \cite{wu2020entropic} and can be regarded as the zero temperature limit of the system since, as we will see in the upcoming sections, when $|A|$ approaches unity, the power in the nonlinear waveguide array system tends to occupy the lowest or the highest-order mode. To explicitly evaluate the sum in Eq. \eqref{eq:2}, we use a polynomial expansion (see Appendix). By employing the global equation of state along with the following identity \cite{weisstein2002binomial,prudnikov1986integrals}
\begin{eqnarray}
\sum^{\infty }_{v=0}{\left( \begin{array}{c}
2v+s \\ 
v \end{array}
\right)}y^v=\frac{2^s}{{\left(1+\sqrt{1-4y}\right)}^s\sqrt{1-4y}} ,
\label{eq:3}
\end{eqnarray}
\noindent one can show that Eq. \eqref{eq:2} can be expressed as follows
\begin{eqnarray}
\frac{1}{\sqrt{1-A^2}}\left[1-\zeta (A,M)\right]=\frac{2\mathcal{P}\kappa }{2\kappa \mathcal{P}-AU}  ,
\label{eq:4}
\end{eqnarray}

\noindent where $\zeta \left(A,M\right)=\frac{2r^2}{M\left(1-r^2\right)}-\frac{2\left(M+1\right)r^{2\left(M+1\right)}}{M(1-r^{2\left(M+1\right)})}$ and $r=\frac{A}{{\left(1+\sqrt{1-A^2}\right)}^{\ }}.$ In order to obtain the optical temperature and chemical potential, one has to first obtain $A\ $from Eq. \eqref{eq:4}. In all cases, we ignore the trivial solution $A=0$ and we only invoke the second root (obtained numerically) which stands for the only physical solution of this problem. Figures 2(a) and 2(b) provide a parametric plot of the variable $A$ as function of the internal energy $U$ and the optical power $\mathcal{P}.$ Note that throughout this paper, for simplicity we set $\kappa =1.$


\begin{figure}
\includegraphics*[width=3in, height=2in]{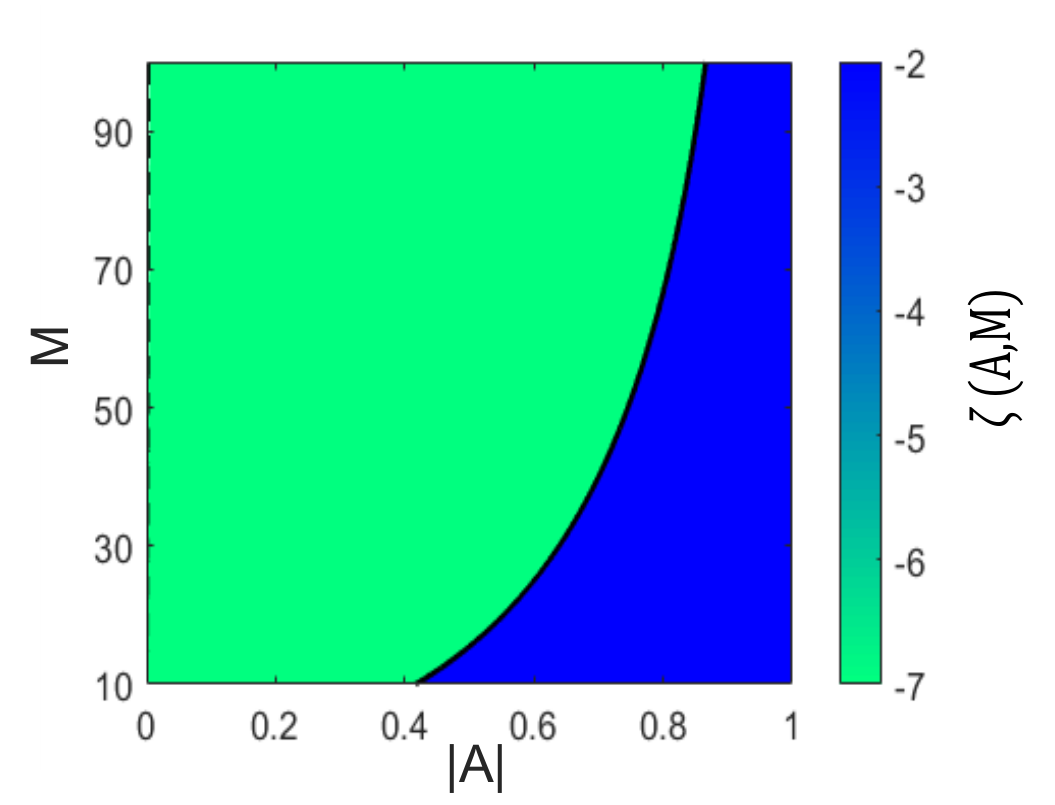}
\caption{\label{fig:3}  A contour plot of $\log_{{10}}{|}{\zeta }\left({A,M}\right){|}$. The solid line indicates the boundary between the regions where $\zeta (A,M)$\textit{  }is above or below ${1\%}$. In this diagram ${M>10}$.}
\end{figure}

From here, the chemical potential $\mu =\frac{2\kappa }{A}$ and the optical temperature $T=\frac{U}{M}-\frac{\mu \mathcal{P}}{M}$ \cite{wu2019thermodynamic} can be subsequently found. It is worth noting that Eq. \eqref{eq:4} is exact in the sense that it is applicable for any temperature $T$. On the other hand, as the number of modes becomes very large ($M\gg 1$) the quantity $A$ can be analytically obtained using the procedure in ref. \cite{wu2020entropic} [$A=4\kappa \mathcal{P}U/(U^2+4{\kappa }^2{\mathcal{P}}^2)]$. By keeping in mind that $A=\frac{2\kappa \mathcal{P}}{U-MT}$ , we get:

\begin{eqnarray}
\frac{1}{\sqrt{1-A^2}}=\frac{2\mathcal{P}\kappa }{2\kappa \mathcal{P}-AU},
\label{eq:5}
\end{eqnarray}

\begin{figure}
\includegraphics*[width=3in, height=6in]{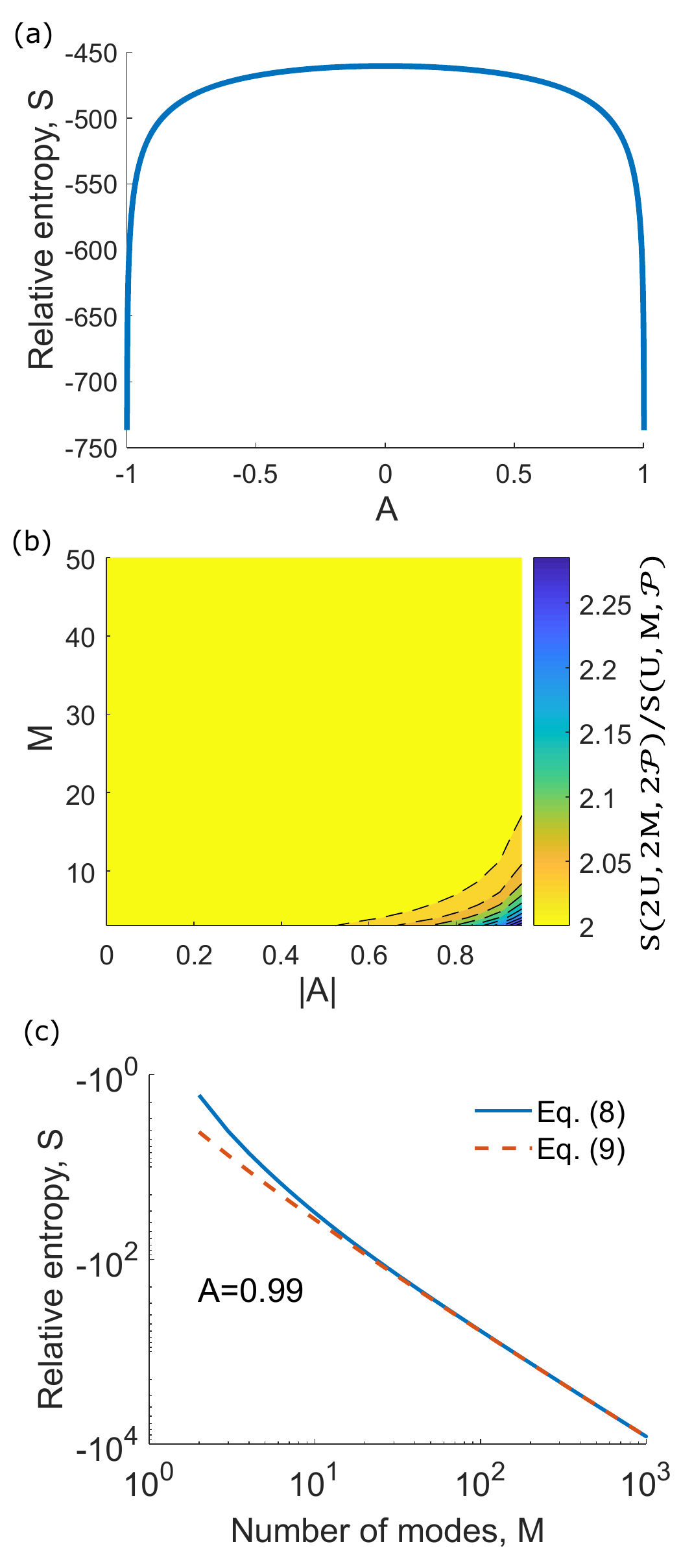}
\caption{\label{fig:4}(a) Optical thermodynamic entropy when $\mathcal{P}{=1\ }$and${\ M=100.}$ The entropy is maximum at ${T}{\to }{\pm }{\infty }$  where ${A=0.}$  (b) A contour plot for the function ${S(2U,2}\mathcal{P}{,2M)/S(U,}\mathcal{P}{,M)}$, where a value of 2 indicates perfect extensivity of the entropy function. Note that the extensivity of this system is preserved for ${M}{\gtrsim }{20}$. (c) A comparison between the entropy provided by Eqs. \eqref{eq:8} and \eqref{eq:9} when  ${A=0.99\ }$and${\ }\mathcal{P}{=1.}$ The two curves coincide with each other for ${M}{\gtrsim }{10}$.}
\end{figure}

\noindent Equation \eqref{eq:4} reduces to Eq. \eqref{eq:5} for highly multimoded systems ($M\gg 1)$ and provided that $A$ lies away from the limit $\left|A\right|=1$, and hence, the function $\zeta (A,M)$ can be ignored. In fact, the $\zeta (A,M)$ term reflects the contribution of discreetness in the kinetic energy $U$ arising at low temperatures whenever the number of modes is very low, an aspect that was unimportant and hence ignored in \cite{wu2020entropic}. Figure 3 depicts the logarithm of the function$\ \zeta (A,M)$, which itself represents a departure from Eq. \eqref{eq:5}. The boundary between the regions where $\zeta (A,M)$ is above or below $1\%$ is also shown for $M>10$.

\noindent According to the second law of thermodynamics, at thermal equilibrium, the entropy of the system should be at maximum. In a photonic system, the thermodynamic entropy can be expressed as follows \cite{wu2019thermodynamic}

\begin{figure}
\includegraphics*[width=3in, height=2in]{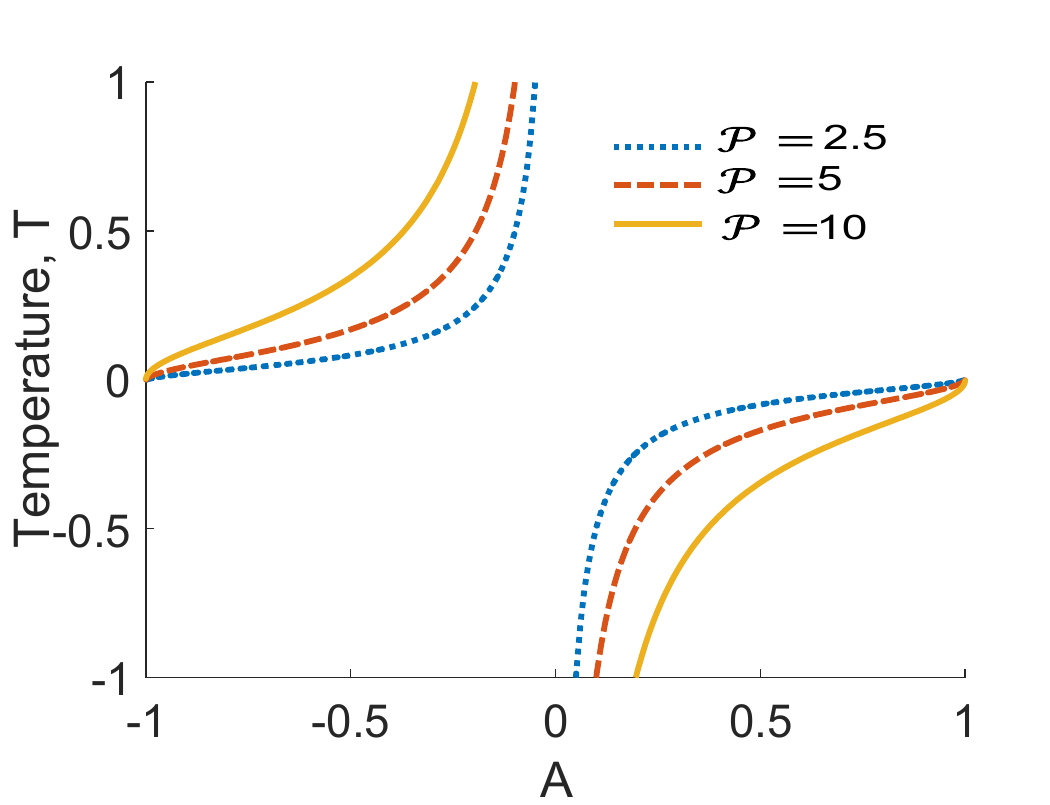}
\caption{\label{fig:5} Optical temperature versus ${A}$. These results are obtained for ${M=100\ }$and ${\kappa }{=1.}$}
\end{figure}

\begin{eqnarray}
S=\sum^M_{j=1}{{\mathrm{ln} {\left|c_j\right|}^2 }}=\sum^M_{j=1}{\mathrm{ln} \left(\frac{-T}{\mu }\right)} \nonumber \\ -\mathrm{ln}\left[A{\mathrm{cos}\left( \frac{j\pi }{M+1}\right)\ }+1\right] .
\label{eq:6}
\end{eqnarray}

\noindent By using the following identity \cite{gradshteyn2014table}
\begin{eqnarray}
\begin{aligned}
&\prod^{N-1}_{k=0}{\ }\left[z^2_1-2z_1z_2\mathrm{cos}\left(\gamma +\frac{2k\pi }{N}\right)+z^2_2\right]\\
&=z^{2N}_1-2z^N_1z^N_2{\mathrm{cos} \left(N\gamma \right)\ }+z^{2N}_2,
\label{eq:7}
\end{aligned}
\end{eqnarray}                                                                                                     

\noindent with $z^2_1=\frac{A^2}{2\left(1+\sqrt{1-A^2}\right)}$ and $z^2_2=\frac{1+\sqrt{1-A^2}}{2}$, the photonic entropy can then be simplified (see Appendix)

\begin{eqnarray}
\begin{aligned}
S= & M\mathrm{ln}\left(\frac{-T}{\mu}\right)-\\
   &\mathrm{ln}\left|\frac{A^{M+1}}{2^{M+1}}{\left(r^{M+1}-r^{-M-1}\right)}\right|\\
   &+\mathrm{ln}\sqrt{1-A^2}.  
\end{aligned}
\label{eq:8}
\end{eqnarray}

\begin{figure}
\includegraphics*[width=3in, height=4in]{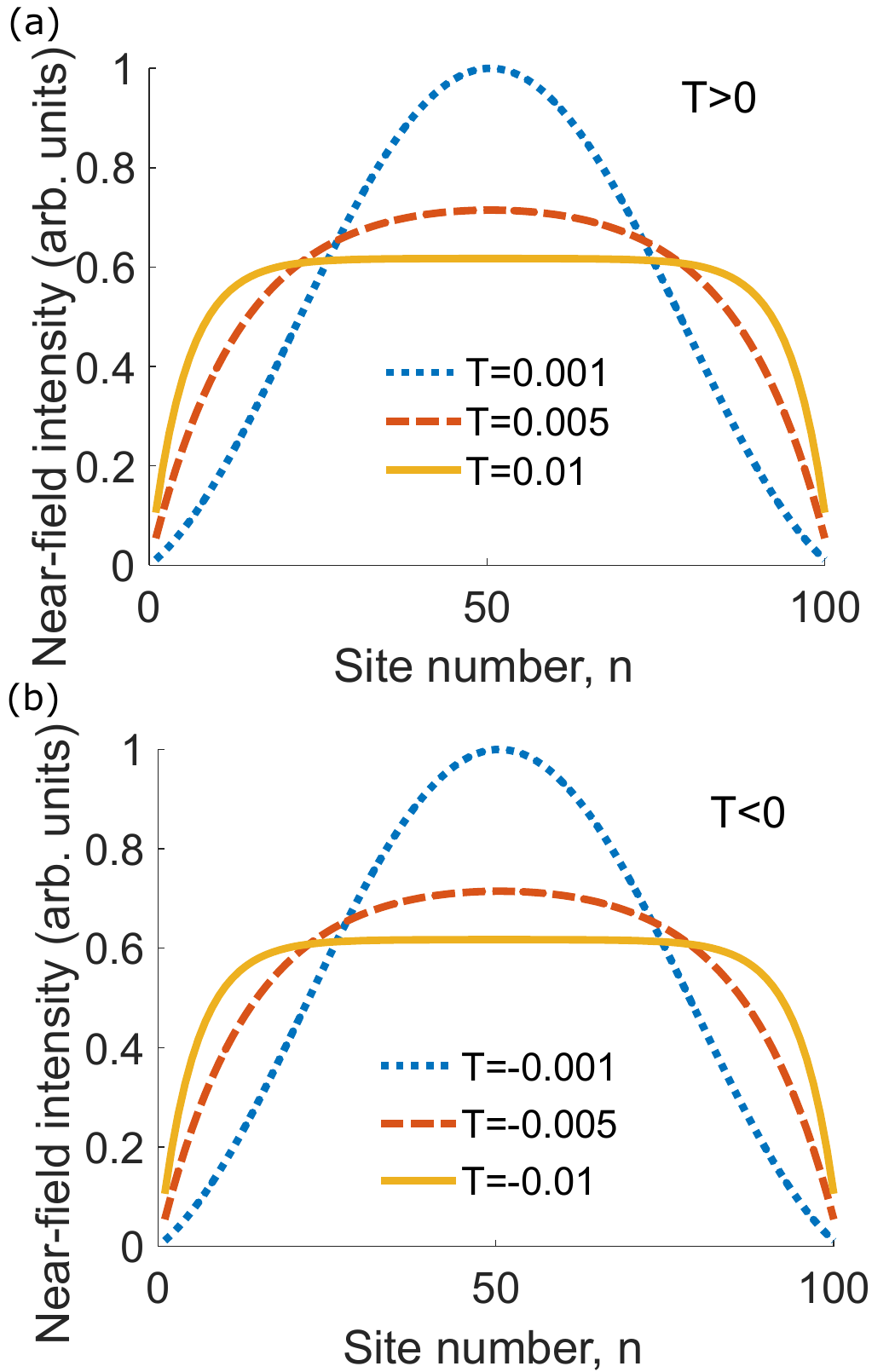}
\caption{\label{fig:6} Theoretical near-field patterns at (a) positive and (b) negative temperatures. Note that the near-field intensity distributions are even functions of temperature. These results were obtained for ${M=100}$, ${\kappa }{=1}$, and $\mathcal{P}{=5.}$  For simplicity, in all the figures displayed in this section, only the envelope of the intensity near-field patterns is provided. In an actual waveguide array system, these patterns are modulated in space by the mode profiles of the waveguide elements involved.}
\end{figure}

 \begin{figure}
\includegraphics*[width=3in, height=2in]{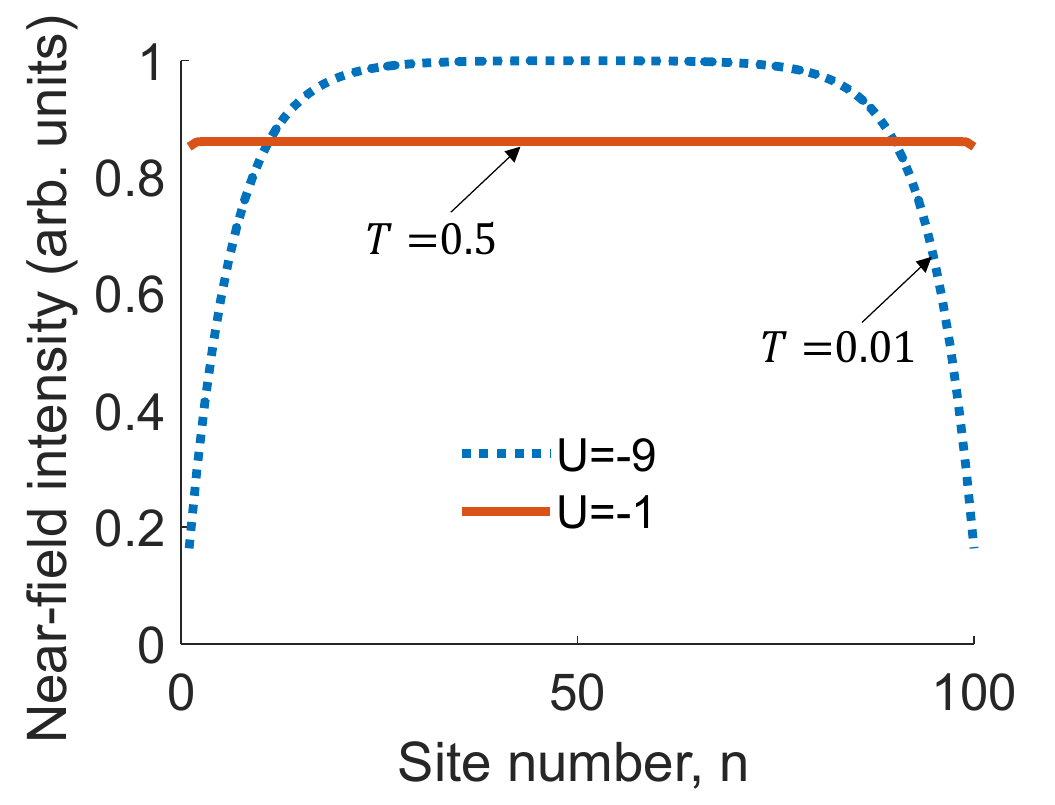}
\caption{\label{fig:7} Near-field intensity distribution at the output of a nonlinear waveguide array as the temperature increases. For these figures$\mathrm{\ }\mathcal{P}{=5}$ and ${M=100}$.}
\end{figure}

 Note that the global equation of state also dictates that  $\frac{T}{\mu }=\frac{AU-2\kappa P}{2M\kappa }$. Given that $A$ can be obtained from Eq. \eqref{eq:4}, the system's entropy can now be readily determined from the initial conditions that uniquely specify $U\ $and $\mathcal{P}$. Fig. 4(a) depicts the dependence of the entropy $S$ on the parameter\textit{ }$A.$ Overall, as $A$ approaches the limits $\mathrm{1}$ or $\mathrm{-}\mathrm{1}$, the photonic entropy $S$ tends to negative infinity while the temperature goes to zero. This zero temperature behavior is very similar to that expected from the Sackur-Tetrode equation describing an ideal photon gas \cite{wu2020entropic}. On the other hand, the entropy is maximum at $A\to 0.$ In this latter limit, the temperature is infinite ($T\to \pm \infty )$  and the system is in its most chaotic state. As a result, all the waveguide modes are equally occupied (power equipartition). At this point, it is worth emphasizing that within the framework of optical thermodynamics one relies on the premise that a very large (yet finite) number of modes is involved. In fact, violation of this very assumption could lead to a scenario where the entropy is not extensive, i.e.,\textit{ }$S\left(\lambda U,\lambda P,\lambda M\right)\neq \lambda S\left(U,P,M\right)$. This aspect is explicitly accounted for by the second term in Eq. \eqref{eq:8} where the extensivity of the entropy is broken as $\left|A\right|\to 1\ $ $\mathrm{and\ }M\to 0$  [see Fig. 4(b)]. On the other hand, if the system is operated away from this extreme regime, Eq. \eqref{eq:6} reduces to optical Sackur-Tetrode equation obtained in ref. \cite{wu2020entropic} that provides the entropy of the photon gas in nonlinear chain systems:

\begin{eqnarray}
\begin{aligned}
 S\left(U,M,\mathcal{P}\right)\approx Mln\left(\frac{4{\kappa }^2{\mathcal{P}}^2-U^2}{4M{\kappa }^2\mathcal{P}}\right). 
\end{aligned}
\label{eq:9}
\end{eqnarray} 

It can be seen that the entropy of Eq. \eqref{eq:9} is extensive in terms of the variables ($U,M,\mathcal{P}$). Figure 4(c) depicts a comparison between the exact expression for the entropy [Eq. \eqref{eq:8}] and the approximate entropy given by Eq. \eqref{eq:9} when $A=0.99.$ As the number of modes increases, the two expressions tend to be asymptotically the same. Interestingly, these two forms are in very close agreement, even though the temperature is very low ($A\to 1$) and the number of modes $M\ $is rather small.

\begin{figure*}
\includegraphics*[width=6in, height=3.5in]{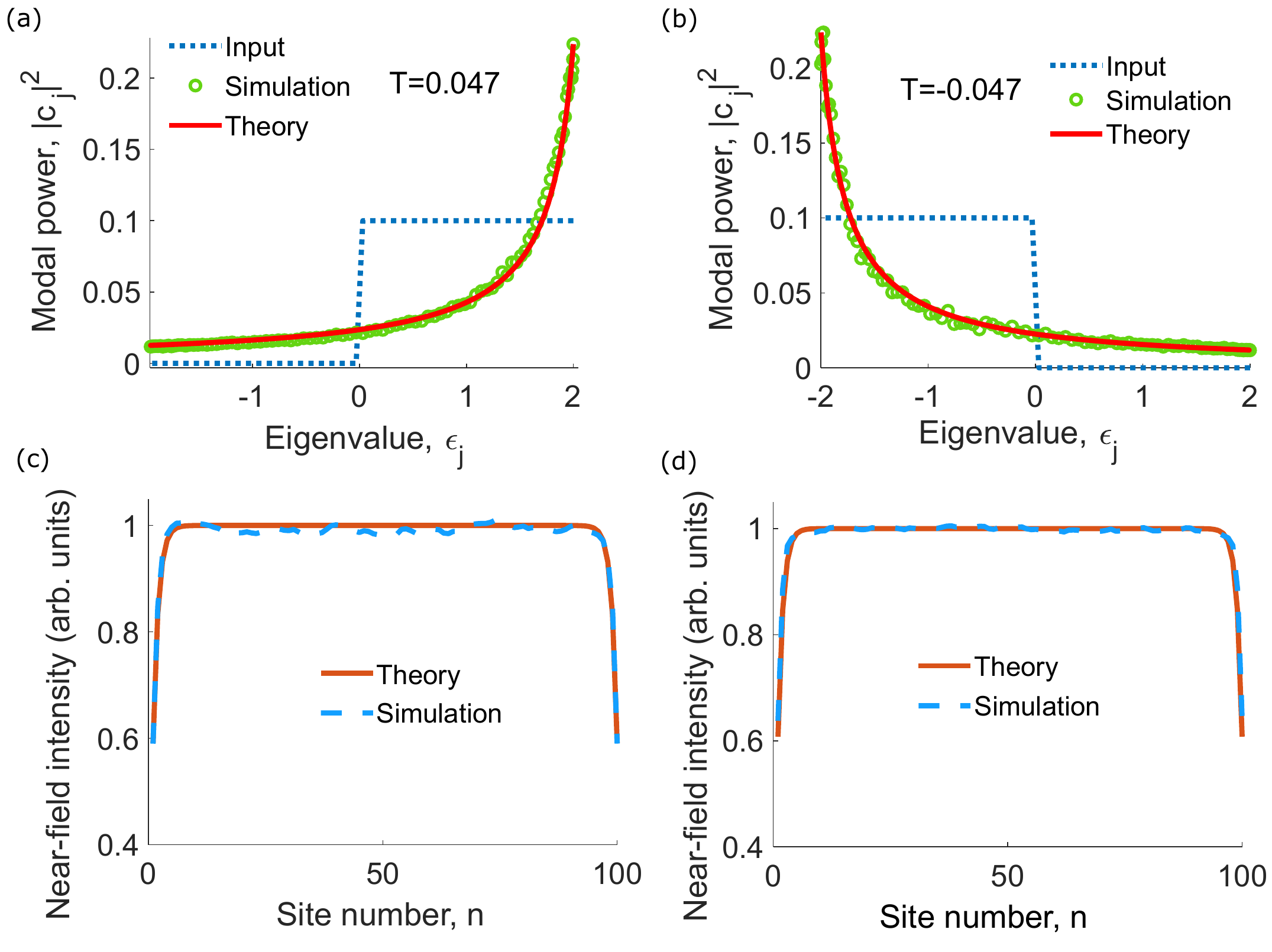}
\caption{\label{fig:8} (a) Resulting modal Rayleigh-Jeans distributions when a nonlinear waveguide array is mostly excited in its lowest-order modes (dotted blue curve), corresponding to the internal energy of  ${U=-6.33.}$ (b) Same as in (a) but with the highest-order modes excited $\left({U=6.33}\right).$ (c) Near-field intensity pattern corresponding to (a). The theoretical intensity distribution expected from Eq. \eqref{eq:13} is compared to direct numerical simulations, as obtained after averaging over several statistical ensembles. (d) Same as in (c). In this case, the near-field intensity pattern corresponds to the initial conditions used in (b). These results were obtained for ${M=100}$, $\mathcal{P}{=5}$, and ${\kappa }{=1.}$}
\end{figure*}

\begin{figure}
\includegraphics*[width=3in, height=3.2in]{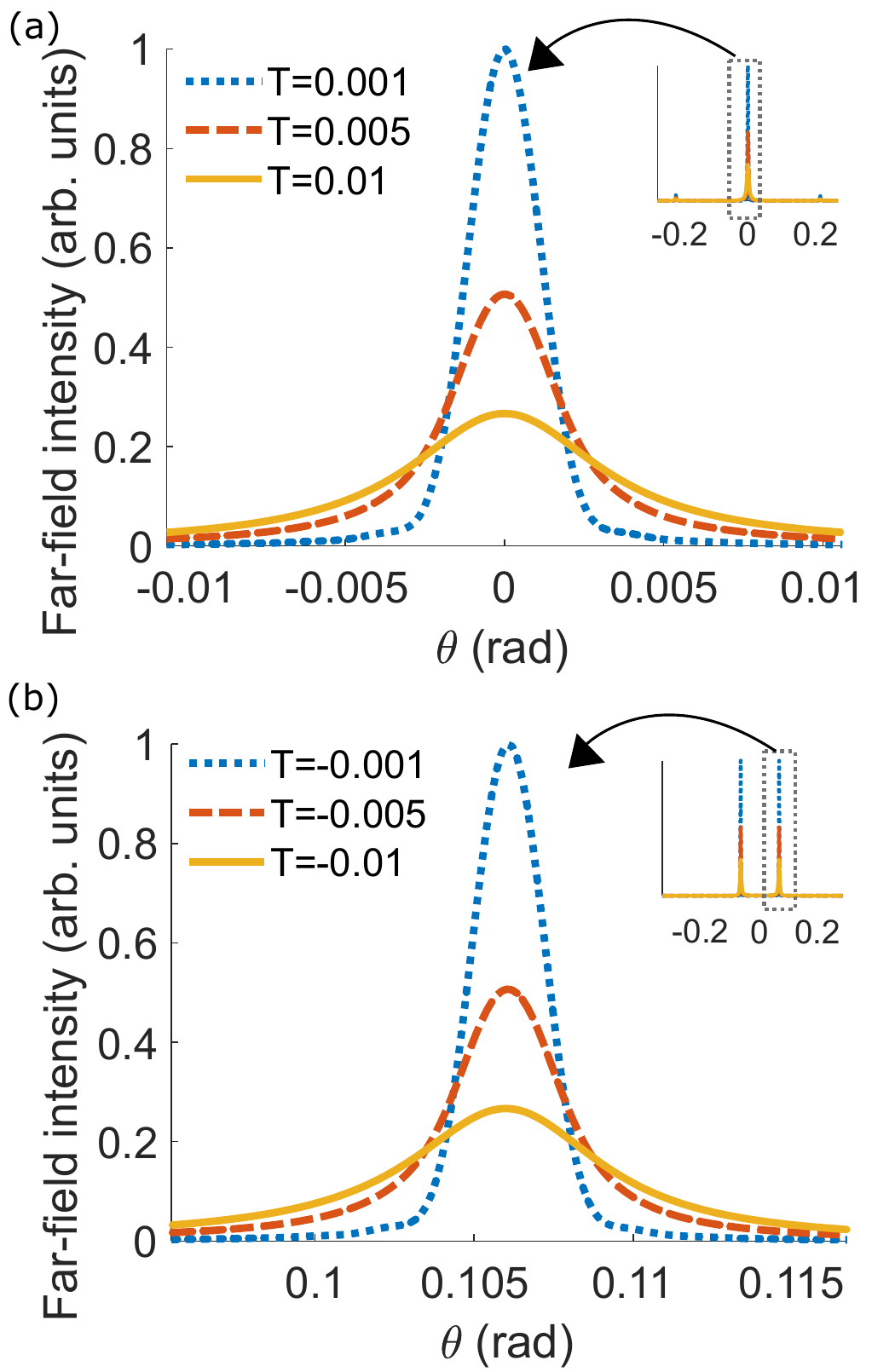}
\caption{\label{fig:9} Far-field intensity patterns as a function of the diffraction angle ${\theta }$ at (a) positive and (b) negative optical temperatures. For the cases considered here ${M=100\ }$and $\mathcal{P}{=5.\ }$ The insets provide the full far-field patterns while a magnification of the main lobe(s) is displayed in the main panels.}
\end{figure}

Next, we derive the equations of the state pertaining to the entropy of the array as given by Eq. \eqref{eq:8}. To do so, we first note that 

\begin{eqnarray}
\begin{aligned}
\frac{\partial S(U,M,\mathcal{P};A)}{\partial U}{\left.{}\right|}_{\mathcal{P},M}=\frac{\partial S}{\partial U}+\frac{\partial S}{\partial A}\frac{\partial A}{\partial U}=\frac{\partial S}{\partial U}, 
\end{aligned}
\label{eq:10}
\end{eqnarray} 

\noindent  given that $\partial S/\partial A=0$ as shown in the Appendix. From here, the temperature $T$, chemical potential $\mu $ and optical pressure $\hat{p}$ can be directly obtained via 

\begin{subequations}
\begin{align}
&{\frac{\partial S}{\partial U}}{\left.{}\right|}_{\mathcal{P},M}
=-\frac{MA}{2\kappa \mathcal{P}-AU}=\frac{1}{T} \quad ,
\label{subeqn:11a} \\
 &\frac{\partial S}{\partial \mathcal{P}}{\left.{}\right|}_{U,M} =\frac{2M\kappa }{2\kappa \mathcal{P}-AU}=-\frac{\mu }{T}\quad , \label{subeqn:11b}\\
&\frac{\hat{p}}{T}={\mathrm{ln} \left(-\frac{AU-2\kappa \mathcal{P}}{2M\kappa e}\right)\ }\nonumber\\& -\frac{\partial }{\partial M}\sum^M_{j=1}{{\mathrm{ln} \left[A{\mathrm{cos} \left(\frac{j\pi }{M+1}\right)\ }+1\right]\quad,}}   \label{subeqn:11c}
\end{align}
\end{subequations}

\noindent where $e$ is Euler's number. Figure 5 shows the relationship between the temperature and the parameter $A$ as obtained from Eq. \eqref{subeqn:11a} for different optical power $\mathcal{P}.$ Note that the first two equations [Eqs. \eqref{subeqn:11a} and \eqref{subeqn:11b}] are equivalent to the global equation of state $U-\mu \mathcal{P}=MT$ as one should expect. In addition, if the system is operated away from zero temperatures and provided that the number of modes is large, Eqs. (11) reduce to Eqs. (2-4) of ref. \cite{wu2020entropic}

\section{Near-field intensity patterns emerging from waveguide arrays  under thermal EQUILIBRiUM CONDITIONS}
We next use the aforementioned thermodynamic description in order to obtain the near-field pattern emerging from a nonlinear waveguide array when it has attained thermal equilibrium at a temperature $T$ and chemical potential $\mu $. In this respect, let us keep in mind that the normalized near-field intensity distribution (among elements $n)$ corresponding to each supermode $\left.|{\psi }_j\right\rangle \ $ is given by   $I_j(n)=\frac{2}{M+1}{\mathrm{sin}}^2(\frac{nj\pi }{M+1})$. In view of the fact that the power distribution among modes is governed by a Rayleigh-Jeans law ${\left|c_j\right|}^2=-T/{(\varepsilon }_j+\mu )$, one can then conclude that the envelope of the thermalized near-field intensity pattern $I_\textrm{NF}(n)$ emerging from the $n$th output site of a nonlinear multimode waveguide array is given by the sum
\begin{equation}
\begin{aligned}
I_\textrm{NF}(n)&=\sum^M_{j=1}{{\left|c_j\right|}^2I_j(n)}\\&=-\frac{2}{M+1}\sum^M_{j=1}{\frac{T/\mu }{A\mathrm{cos}\left(\frac{j\pi }{M+1}\right)+1}\mathrm{si}{\mathrm{n}}^2\left(\frac{nj\pi }{M+1}\right)}.
\label{eq:12}
\end{aligned}
\end{equation} 

\noindent Using the sum identities outlined in the Appendix, and provided that the array is heavily multimoded ($M\gg 1)$ and is operated away from zero temperatures $(\left|A\right|\neq $1), one can show that the near-field intensity profile can be approximately expressed as follows
\begin{equation}
\begin{aligned}
I_\textrm{NF}(n)\approx \left\{ \begin{array}{c}
\frac{-T}{\mu\sqrt{1-A^2}}\left(1-\frac{Mr^{2n}}{M+1}\right)\ 1\le n\le L \\ 
\frac{-T}{\mu\sqrt{1-A^2}}\left(1-\frac{Mr^{2M+2-2n}}{M+1}\right)L+1\le n\le M.\end{array}
\right.
\label{eq:13}
\end{aligned}
\end{equation}

 \noindent where $L$ is the closest integer less than or equal to $M/2.$ The near-field intensity patterns as obtained from Eq. \eqref{eq:12} are depicted in Fig. 6, under different temperature conditions when $\mathcal{P}\mathrm{=5}$. In general, as the magnitude of the optical temperature $|T|$ increases, the entropy is larger and as a result the beam intensity pattern tends to be flatter. On the other hand, when the temperature approaches $0^{\pm }$, the intensity profile is solely represented by the fundamental or the highest-order mode. Interestingly, regardless of the sign of the temperature, as long as the entropy is the same, the system exhibits the same near-field pattern. This is due to the fact that optical entropy $S$ is an even function of temperature $T$ in 1-D waveguide array systems.

We next show that the near-field distribution can be controlled at will by judiciously varying the initial excitation conditions. Figure 7 shows that for a given input power ($\mathcal{P}\mathrm{=5}$), the output intensity profile tends to become relatively constant across the array as the internal energy $U$ approaches zero, in which case the temperature tends to infinity. This is because as $T\to \pm \infty $ equipartition of power takes place among modes and therefore the intensity profile flattens out.

To corroborate the aforementioned thermodynamic results, we performed numerical simulations based on direct numerical integration of Eq. \eqref{eq:1}. In these simulations  $M=100$\textit{,} $\kappa =1$ and $\mathcal{P}=5$. Figures 8(a) and 8(b) depict the equilibrium Rayleigh-Jeans distributions when the array system is excited with internal energies $U=\mp 6.33$. In the first scenario [Fig. 8(a)], the lowest order modes (dotted blue curve) are excited while in the second the power is injected in higher-order states [Fig. 8(b)]. In both cases, there is excellent agreement between simulations and the thermodynamic formalism used above. Under these conditions, the system attains thermal equilibrium at a temperature of   $T=\pm 0.047$, corresponding to $U=\mp 6.33$. In all cases, the resulting site intensity distributions$\ I_\textrm{NF}(n)$ across the array are obtained by numerically solving Eq. \eqref{eq:1}, and after averaging the results of several ensembles, all initiated with the same modal amplitudes ${\left|c_{j0}\right|}^{\ }$ but with different random phases $[\mathrm{arg}\mathrm{}(c_{j0})]$. These results are then compared to the intensity distribution $I_\textrm{NF}(n)$ predicted by Eq. \eqref{eq:13}. As indicated in Figs. 8(c) and 8(d), in both cases the near-field intensity patterns obtained from numerical simulations are in very close agreement with theoretical predictions. The small fluctuations appearing in Fig. 8(c) are due to weak modulational instability effects that tend to affect lower-order in-phase modes under self-focusing nonlinearities, which is the case examined in this paper. 
\section{Array Far-field intensity patterns under thermal EQUILIBRiUM CONDITIONS}
In this section, we investigate the far-field intensity patterns emitted by a nonlinear waveguide array system once it attains thermal equilibrium. As opposed to near-field intensity distributions, the far-field patterns can be obtained by invoking diffraction effects. Given that the relative phases among modes vary in a stochastic manner, the far-field intensity can be determined by incoherently superimposing the diffraction pattern of each array supermode.  In this case, the electric field distribution of each mode can be obtained by convolving the array supermode with the mode profile of each single-mode waveguide channel $g\left(x\right)\ $(the local mode) \cite{berger1986far}. Therefore, the electric field distribution for each mode $j\ $can be written as 
\begin{figure*}
\includegraphics*[width=6in, height=3.8in]{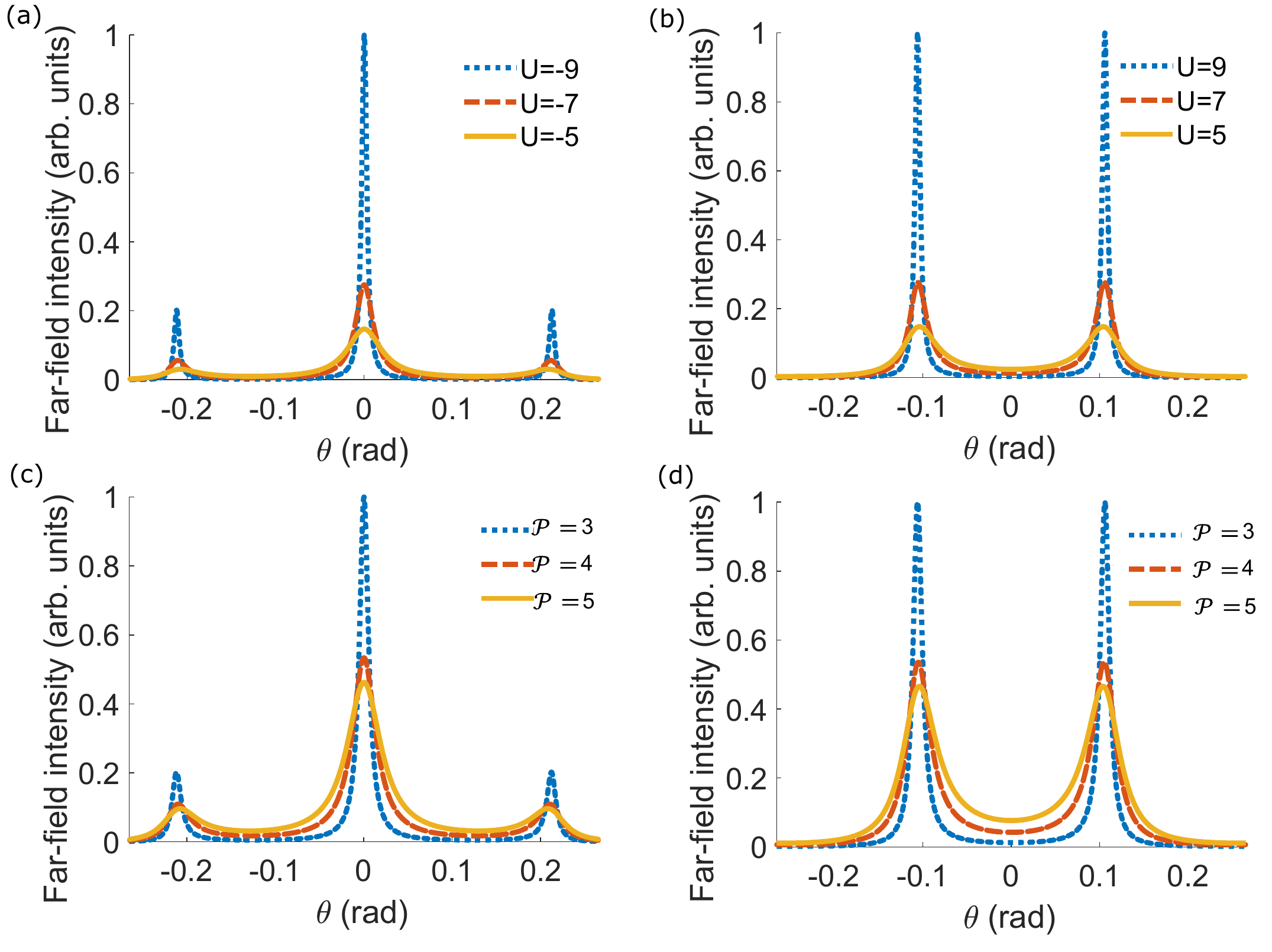}
\caption{\label{fig:10}  Far-field intensity patterns as a function of internal energy ${U}$ at (a) positive and (b) negative temperatures when $\mathcal{P}{=5}$. (c) and (d) same as in (a) and (b) with the power varying while ${U}$ is equal to ${-}{5\ }$and ${5}$, respectively.}
\end{figure*}

\begin{equation}
\begin{aligned}
E_j\left(x\right)=c_j\sqrt{\frac{2}{M+1}}\sum^M_{n=1}{\ \int{\ d\eta }}\mathrm{sin}\left(\frac{nj\pi }{M+1}\right) \\ \times \delta \left(\eta -nD\right)g\left(x-\eta \right)\  
\label{eq:14}
\end{aligned}
\end{equation}

\noindent where\textit{ D} is the waveguide separation distance and $x$ is the transverse coordinate in the waveguide array plane. Here, for simplicity, we assume that the modal field profile is Gaussian with a spot size $w$, i.e., $g\left(x\right)=\mathrm{exp}\mathrm{}(-x^2/w^2)$. By adopting, the results of ref. \cite{butler1984coupled}, the Fraunhofer far-field intensity pattern $I_{\textrm{FF}}(\theta )$ resulting from incoherent superposition of all the supermodes is given by:

\begin{widetext}
\begin{equation}
 I_{\mathrm{FF}}(\theta )=f\left(\theta \right)\sum^M_{j=1}{{\left|c_j\right|}^2\times {\left\{\ \frac{{\mathrm{sin} \left[\left({\phi }_j+\alpha \theta \right)\frac{M}{2}\ \right]\ }} {{\mathrm{sin} \left(\frac{{\phi }_j}{2}+\frac{\alpha \theta }{2}\right)\ }}-\frac{{\left(-1\right)}^j{\mathrm{sin} \left[\left(\alpha \theta -{\phi }_j\right)\frac{M}{2}\right]\ }}{{\mathrm{sin} \left(\frac{\alpha \theta }{2}-\frac{{\phi }_j}{2}\right)\ }}\right\}}^2} 
 \label{eq:15}
\end{equation}
\end{widetext}

\noindent where $\theta $ is the far-field diffraction angle, $\alpha =\frac{2\pi }{{\lambda }_o}D$,  ${\phi }_j=\frac{j\pi }{M+1}$. Since the mode in each waveguide is Gaussian, $f\left(\theta \right)\propto \mathrm{exp}\mathrm{}(-2{\theta }^2/{\theta }^2_o)$ where, ${\theta }_o={\lambda }_o/(\pi w).$ Under thermal equilibrium conditions, the far-field intensity distribution emitted by a nonlinear waveguide array is dictated by Eq. \eqref{eq:15} provided that the modal occupancies ${\left|c_j\right|}^2$ obey the Rayleigh-Jeans law. As an example, let us consider a nonlinear waveguide array operated at a wavelength of ${\lambda }_o=1\ \mu m.\ $ The separation among waveguides is $4.7 \ \mu m$ and the mode spot size of each element is $w\approx 1.9\ \mu m.$ As indicated by Figs. 9(a) and 9(b), the far-field intensity patterns directly depend on the optical temperature of this system at equilibrium.  It is worth emphasizing that as the magnitude of the optical temperature increases, the beam divergence in the far-field also increases. At positive temperatures, the far-field pattern is dominated by a central lobe [Fig. 9(a)] since lower-order modes are favored in the array. Conversely, at negative temperatures, the far-field is split into two lobes [Fig. 9(b)], as expected from higher-order modal states.

The dependence of the far-field intensity on the internal energy $U\ $is shown in Figs. 10(a) and 10(b) when the power level is kept constant $\mathcal{P}=5\ .$ As these figures suggest, the far-field lobes tend to broaden as the magnitude of internal energy $|U|$ decreases. This is because the system progressively moves towards higher absolute temperatures-thus spoiling the spatial coherence of the light emitted. Similarly, Figs. 10(c) and 10(d) depict how the same patterns vary as the power $\mathcal{P}$ in the array increases while keeping $U $ constant. In this case, the pattern broadens at high powers, in both temperature regimes. 

\noindent

In order to validate the predictions of the optical thermodynamic theory, the theoretical results in this section are compared to those obtained numerically after integrating Eq. \eqref{eq:1}. To do so, as in section III, statistical ensembles have been used. Figure 11 depicts the far-field patterns emerging from a nonlinear waveguide array at two different excitation conditions. Figure 11(a) compares the numerical simulations (after summing incoherently the Fraunhofer patterns of each spatial supermode or by summing the Fraunhofer diffraction patterns of near-field intensity profile over many ensembles) with our analytical results [Eq. \eqref{eq:15}]. The excitation conditions used are those corresponding to  Fig. 8(a). In the same vein, Fig. 11(b) compares these simulations with the results predicted by Eq. \eqref{eq:15}, under the same conditions used in Fig. 8(b). In all cases, good agreement was found between the numerical and theoretical results.

\begin{figure}
\includegraphics*[width=3in, height=3.8in]{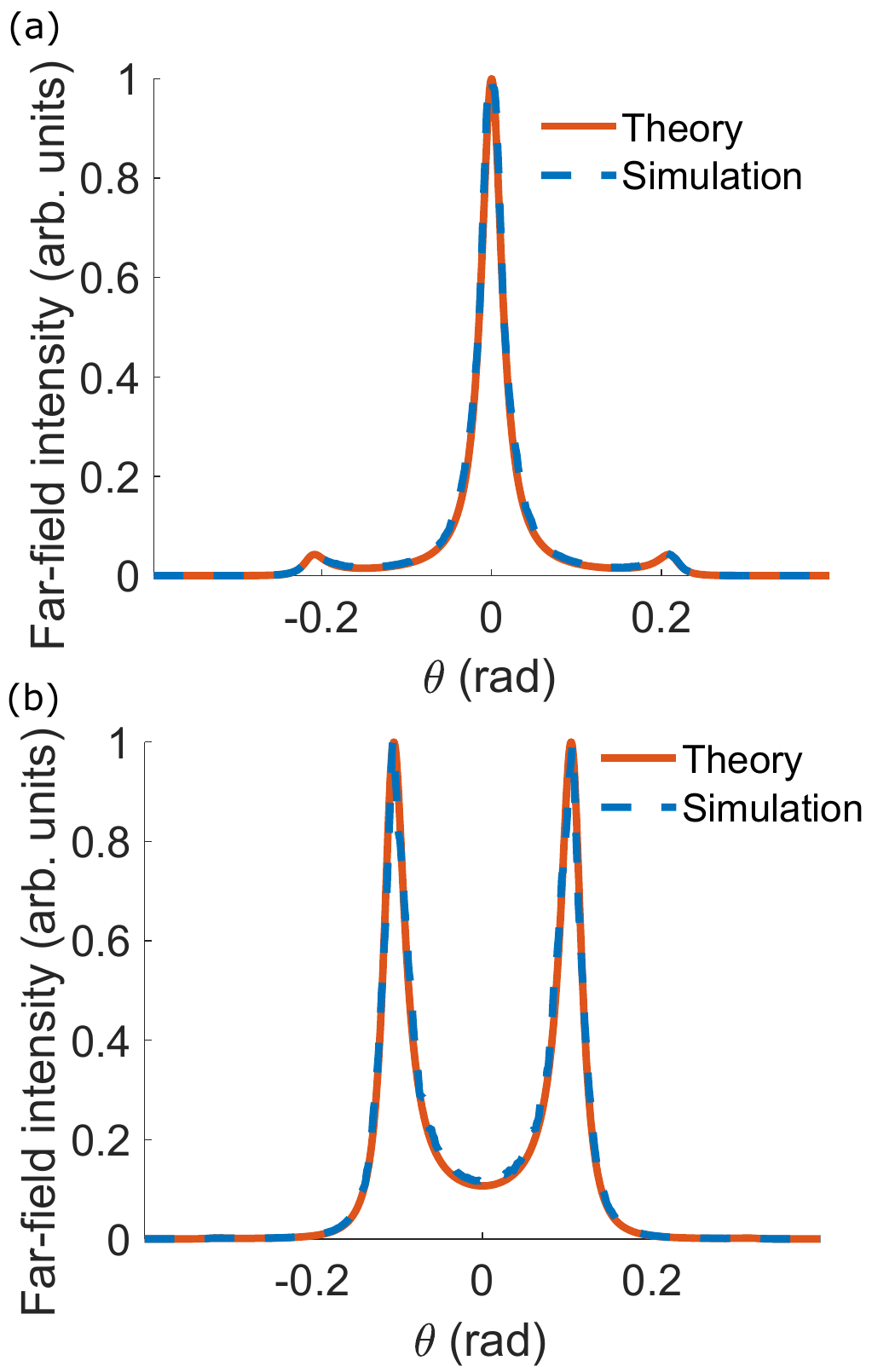}
\caption{\label{fig:11} Numerically validating the far-field intensity patterns, as predicted by the optical thermodynamic theory. Theoretical and numerical results are plotted together for (a) ${T}{\approx }{+0.047}$ and (b) ${T}{\approx }{-0.047}$. These results were obtained for ${M=100}$, ${U=}{\mp }{6.33}$, $\mathcal{P}{=5,}$ ${\kappa }{=1}$, and ${D=4.7}{\ }{\mu }{m.}$}
\end{figure}

\section{conclusion}
In conclusion, by means of optical thermodynamics, we provided a systematic approach for analytically predicting and describing the near- and far-field radiation patterns emerging from waveguide chain networks. In general, we found that the temperature and chemical potential govern the near- and far-field intensity distributions. Issues related to the extensivity of the system's entropy have also been addressed.  In all cases, good agreement between the analytical results obtained from the optical thermodynamic theory and the numerical simulations has been demonstrated. 
\vspace*{-0.35cm}

\section*{ACKNOWLEDGMENTS}
This work was partially supported by ONR MURI (N00014-20-1-2789), AFOSR MURI (FA9550-20-1-0322, FA9550-21-1-0202), DARPA (D18AP00058), Office of Naval Research (N00014-16-1-2640, N00014-18-1-2347, N00014-19-1-2052, N00014-20-1-2522, N00014-20-1-2789), National Science Foundation (NSF) (DMR-1420620, EECS-1711230, CBET 1805200, ECCS 2000538, ECCS 2011171), Air Force Office of Scientific Research (FA9550-14-1-0037,  FA9550-20-1-0322, FA9550-21-1-0202), MPS Simons collaboration (Simons grant 733682), W. M. Keck Foundation, USIsrael Binational Science Foundation (BSF: 2016381), US Air Force Research Laboratory (FA86511820019) and the Qatar National Research Fund (grant NPRP13S0121-200126).
\vspace*{-0.3cm}
\section*{Appendix}

 We here derive Eq. \eqref{eq:4} from Eq. \eqref{eq:2}.  By using a Taylor series expansion, Eq. \eqref{eq:2} can be written as
\renewcommand{\theequation}{A.\arabic{equation}}

\setcounter{equation}{0}

\begin{equation}
\mathcal{P}=-\frac{T}{\mu }\sum^M_{j=1}{\left[1-A\mathrm{cos}\left({\phi }_j\right)+A^2{{\mathrm{cos}}^{\mathrm{2}} \left({\phi }_j\right)\ }-\dots \right]}, 
\label{eq:A1}
\end{equation}

\noindent where ${\phi }_j=\frac{j\pi }{M+1}\mathrm{.\ }$By using the identities \cite{beyer1991crc}

\noindent 

\begin{eqnarray}
&& {{\mathrm{cos}}^{2v} a\ }=\frac{1}{2^{2v}}\left( \begin{array}{c}
2v \\ 
v \end{array}
\right) \nonumber \\ && +\frac{1}{2^{2v-1}}\sum^{v-1}_{k=0}{\ }\left( \begin{array}{c}
2v \\ 
k \end{array}
\right)\mathrm{cos}\mathrm{}\mathrm{[}2\left(v-k\right)a\mathrm{]}, \label{eq:A2}\\ && {{\mathrm{cos}}^{2v+1} b\ }=\frac{1}{4^v}\sum^v_{k=0}{\ }\left( \begin{array}{c}
2v+1 \\ 
k \end{array}
\right)\nonumber \\ && \qquad\qquad\quad   \times \mathrm{cos}\mathrm{}[\left(2v+1-2k\right)b],
\label{eq:A3}
\end{eqnarray}

\noindent with $\left( \begin{array}{c}
2v \\ 
k \end{array}
\right)$ representing the binomial coefficient (i.e., $\left( \begin{array}{c}
2v \\ 
k \end{array}
\right)=2v!/[k!\left(2v-k\right)!)$], we can then expand Eq. \eqref{eq:A1}  as 

\begin{eqnarray}
 && \mathcal{P}=\frac{-T}{\mu }\sum^M_{j=1}{\left[1 +\frac{1}{2^2}\left( \begin{array}{c}
2 \\ 
1 \end{array}
\right)A^2\right.} \nonumber  \\ && \left.+\frac{1}{2^4}\left( \begin{array}{c}
4 \\ 
2 \end{array}
\right)A^4 +\frac{1}{2^6}\left( \begin{array}{c}
6 \\ 
3 \end{array}
\right)A^6+\dots \right]\nonumber \\ && -\left[\frac{A}{4^0}\left( \begin{array}{c}
1 \\ 
0 \end{array}
\right)+\frac{A^3}{4^1}\left( \begin{array}{c}
3 \\ 
1 \end{array}
\right)+\frac{A^6}{4^2}\left( \begin{array}{c}
5 \\ 
2 \end{array}
\right) +\dots \right]{\mathrm{cos} \left({\phi }_j\right)\ } \nonumber \\&&+\left[\frac{A^2}{2^1}\left( \begin{array}{c}
2 \\ 
0 \end{array}
\right)+\frac{A^4}{2^3}\left( \begin{array}{c}
4 \\ 
1 \end{array}
\right)+\frac{A^6}{2^5}\left( \begin{array}{c}
6 \\ 
2 \end{array}
\right)+\dots \right]{\cos \left(2{\phi }_j\right)\ }\nonumber \\ &&-\dots.
\label{eq:A4}
\end{eqnarray}

Now, let us rewrite the term $\sum^M_{j=1}{{\cos \left(\frac{mj\pi }{M+1}\right)\ }}$  using Euler's formula

\noindent 

\begin{eqnarray}
\sum^M_{j=1}{{\mathrm{cos} \left(\frac{mj\pi }{M+1}\right)\ }}=\frac{1}{2}\sum^M_{j=1}{e^{i\frac{mj\pi }{M+1}}+e^{-i\frac{mj\pi }{M+1}}}.
\label{eq:A5}
\end{eqnarray}

\noindent The geometric series identity $\sum^M_{j=1}{b^j}=b\left(1-b^M\right)/(1-b)$ leads to

\noindent 

\begin{eqnarray}
&& \sum^M_{j=1}{{\mathrm{cos} \left(\frac{mj\pi }{M+1}\right)\ }}=\nonumber \\ && \frac{1}{2}\left(\frac{e^{i\frac{m\pi }{M+1}}-e^{im\pi }}{1-e^{i\frac{m\pi }{M+1}}}+\frac{e^{-i\frac{m\pi }{M+1}}-e^{-im\pi }}{1-e^{-i\frac{m\pi }{M+1}}}\right),
\label{eq:A6}
\end{eqnarray}

\noindent which can be further simplified using $\sum^M_{j=1}{{\cos \left(\frac{mj\pi }{M+1}\right)\ }}=[-1$,$0]$ if $m$ is [even, odd], respectively as long as $m\ $ is not an even multiple of $\left(M+1\right).$ If on the other hand, $m=2q\left(M+1\right)$ then $\sum^M_{j=1}{{\cos \left(\frac{mj\pi }{M+1}\right)\ }}=M.$ Hence, Eq. \eqref{eq:A4} is rewritten as 

\noindent 

\begin{eqnarray}
&&\mathcal{P}=-\frac{T}{\mu }\sum^{\infty }_{v=0}{\frac{M}{2^{2v}}\left( \begin{array}{c}
2v \\ 
v \end{array}
\right)A^{2v}-}\frac{A^{2\left(v+1\right)}}{2^{2v+1}}\left( \begin{array}{c}
2v+2\\ 
v \end{array}
\right) \nonumber \\ &&-\frac{A^{2\left(v+2\right)}}{2^{2v+3}}\left( \begin{array}{c}
2v+4 \\ 
v \end{array}
\right)-\dots \nonumber\\ && +M\left[\frac{A^{2v+2M+2}}{2^{2v+2M+1}}\left( \begin{array}{c}
2v+2M+2 \\ 
v \end{array}
\right)\right]+\dots .
\label{eq:A7}
\end{eqnarray}

\noindent By directly substituting Eq. \eqref{eq:3} into Eq. \eqref{eq:A7} we find

\noindent 

\begin{eqnarray}
& \mathcal{P}=-\frac{T}{\mu }\left(\frac{1}{\sqrt{1-A^2}}\right)\left\{M-2\left[r^2+r^4+r^6+\dots \right]\right.
\nonumber \\ &\left.+2(M+1)\left[r^{2(M+1)}+r^{4(M+1)}+r^{6(M+1)}+\dots \right] \right\},\nonumber\\
\label{eq:A8}
\end{eqnarray}

\noindent where $r=\frac{A}{{\left(1+\sqrt{1-A^2}\right)}^{\ }}.$ By reorganizing the geometric series in Eq. \eqref{eq:A8} and by using the relations, $\mu =\frac{2\kappa }{A}$ and $\frac{T}{\mu }=\frac{AU-2\kappa P}{2M\kappa }$, one can obtain Eq. \eqref{eq:4}. 

Next, we derive Eq. \eqref{eq:8} from Eq. \eqref{eq:6}. To this end, Eq. \eqref{eq:6} is rewritten as

\begin{eqnarray}
S=&&M{\mathrm{ln} \left(\frac{-T}{\mu }\right)\ }\nonumber \\ &&-{\mathrm{ln} \prod^M_{j=1}{\left[A{\cos \left(\frac{j\pi }{M+1}\right)\ }+1\right]}\ },
\label{eq:A9}
\end{eqnarray}


\noindent By utilizing the symmetry of the cosine function, we can get 
\begin{eqnarray}
&&\prod^M_{j=0}{\left[A{\cos \left(\frac{j\pi }{M+1}\right)\ }+1\right]}=\sqrt{ \frac{1}{1-A}} \nonumber \\ &&\times \sqrt{\prod^{M+1}_{j=0}{\left[A{\mathrm{cos} \left(\frac{2j\pi }{M+1}\right)\ }+1\right]}}\nonumber \\ &&\times\sqrt{\prod^{M+1}_{j=1}{\left[A{\mathrm{cos} \left(\frac{(2j-1)\pi }{M+1}\right)\ }+1\right]}}.
\label{eq:A10}
\end{eqnarray}

\noindent 

\noindent By using Eq. \eqref{eq:7} and after setting $z^2_1=\frac{A^2}{2\left(1+\sqrt{1-A^2}\right)}$, $z^2_2=\frac{1+\sqrt{1-A^2}}{2}$, one can obtain Eq. \eqref{eq:8}. In the process, $\gamma =0$ and $\gamma =\frac{-\pi }{M+1}$ are used in the second and third product on the right-hand side of Eq. \eqref{eq:A10}, respectively.

We next derive Eq. \eqref{eq:13}. By employing the trigonometric identity ${{\mathrm{sin}}^{\mathrm{2}} p\ }\mathrm{=[}1-\mathrm{cos}\mathrm{}(2p)]/2$\textit{,} Eq. \eqref{eq:12} can be rewritten as
\begin{eqnarray}
  I_\textrm{NF}(n)=&&-\frac{1}{M+1}\sum^M_{j=1}{\frac{\frac{T}{\mu \ }}{A\mathrm{cos}\left(\frac{j\pi }{M+1}\right)+1}}\nonumber \\ && \times\left[1-\mathrm{cos}\left(\frac{2nj\pi }{M+1}\right)\right].
\label{eq:A12}
\end{eqnarray}

\noindent The first term [$\sum^M_{j=1}{\frac{T/\mu \ }{A\mathrm{cos}\left(j\pi /(M+1)\right)+1}}\mathrm{]}$ can be recast in the same way Eq. \eqref{eq:4} was derived. For $M\gg 1$ and $\left|A\right|\neq 1$, we find that $\frac{1}{M+1}\sum^M_{j=1}{\frac{\frac{T}{\mu }}{A\mathrm{cos}\left(\frac{j\pi }{M+1}\right)+1}\approx \frac{T}{\mu }\frac{1}{\sqrt{1-A^2}}}$. This strategy can also be used in evaluating the second term in Eq. \eqref{eq:A12} by employing the  discrete Fourier transform identity

\noindent 

\begin{eqnarray}
&&\sum^M_{j=1}{\left(e^{i2n\frac{j\pi }{M+1}}+e^{-i2n\frac{j\pi }{M+1}}\right){\mathrm{cos} \left(2m\phi_{j}\right)\ }}\nonumber\\&&\approx M\left[\delta (2n-2m)+\delta (2n-2M-2m+2)\right]
\label{eq:A13}
\end{eqnarray}

\noindent From here, one can obtain Eq. \eqref{eq:13}.

Finally, we show that $\frac{\partial S}{\partial A}=0$. By directly substituting $\frac{T}{\mu }=(AU-2\kappa \mathcal{P})/2M\kappa $ into Eq. \eqref{eq:6} and after differentiating both sides with respect to $A$, Eq. \eqref{eq:6} reads

\begin{eqnarray}
 \frac{\partial S}{\partial A}=&&\frac{UM}{AU-2\kappa P} \nonumber \\ &&-\sum^M_{j=1}{\frac{{\mathrm{cos} \left(\frac{j\pi }{M+1}\right)\ }}{\left[A{\mathrm{cos} \left(\frac{j\pi }{M+1}\right)\ }+1\right]}}.
\label{eq:A14}
\end{eqnarray}

\noindent Upon substituting $U=-\sum^M_{j=1}{{\varepsilon }_j}{\left|c_j\right|}^2=\sum^M_{j=1}{\frac{2\kappa {\mathrm{cos} \left(\frac{j\pi }{M+1}\right)\ }T}{2\kappa {\mathrm{cos} \left(\frac{j\pi }{M+1}\right)\ }+\mu }}$ into the latter expression, one finds that

\noindent

\begin{eqnarray}
  \frac{\partial S}{\partial A}=\frac{UM}{AU-2\kappa P}-\frac{UM}{AU-2\kappa \mathcal{P}}=0.
 \label{eq:A15}
\end{eqnarray}




\providecommand{\noopsort}[1]{}\providecommand{\singleletter}[1]{#1}%
\begin{thebibliography}{45}%
\makeatletter
\providecommand \@ifxundefined [1]{%
 \@ifx{#1\undefined}
}%
\providecommand \@ifnum [1]{%
 \ifnum #1\expandafter \@firstoftwo
 \else \expandafter \@secondoftwo
 \fi
}%
\providecommand \@ifx [1]{%
 \ifx #1\expandafter \@firstoftwo
 \else \expandafter \@secondoftwo
 \fi
}%
\providecommand \natexlab [1]{#1}%
\providecommand \enquote  [1]{``#1''}%
\providecommand \bibnamefont  [1]{#1}%
\providecommand \bibfnamefont [1]{#1}%
\providecommand \citenamefont [1]{#1}%
\providecommand \href@noop [0]{\@secondoftwo}%
\providecommand \href [0]{\begingroup \@sanitize@url \@href}%
\providecommand \@href[1]{\@@startlink{#1}\@@href}%
\providecommand \@@href[1]{\endgroup#1\@@endlink}%
\providecommand \@sanitize@url [0]{\catcode `\\12\catcode `\$12\catcode
  `\&12\catcode `\#12\catcode `\^12\catcode `\_12\catcode `\%12\relax}%
\providecommand \@@startlink[1]{}%
\providecommand \@@endlink[0]{}%
\providecommand \url  [0]{\begingroup\@sanitize@url \@url }%
\providecommand \@url [1]{\endgroup\@href {#1}{\urlprefix }}%
\providecommand \urlprefix  [0]{URL }%
\providecommand \Eprint [0]{\href }%
\providecommand \doibase [0]{https://doi.org/}%
\providecommand \selectlanguage [0]{\@gobble}%
\providecommand \bibinfo  [0]{\@secondoftwo}%
\providecommand \bibfield  [0]{\@secondoftwo}%
\providecommand \translation [1]{[#1]}%
\providecommand \BibitemOpen [0]{}%
\providecommand \bibitemStop [0]{}%
\providecommand \bibitemNoStop [0]{.\EOS\space}%
\providecommand \EOS [0]{\spacefactor3000\relax}%
\providecommand \BibitemShut  [1]{\csname bibitem#1\endcsname}%
\let\auto@bib@innerbib\@empty
\bibitem [{\citenamefont {Renninger}\ and\ \citenamefont
  {Wise}(2013)}]{renninger2013optical}%
  \BibitemOpen
  \bibfield  {author} {\bibinfo {author} {\bibfnamefont {W.~H.}\ \bibnamefont
  {Renninger}}\ and\ \bibinfo {author} {\bibfnamefont {F.~W.}\ \bibnamefont
  {Wise}},\ }\bibfield  {title} {\bibinfo {title} {Optical solitons in
  graded-index multimode fibres},\ }\href@noop {} {\bibfield  {journal}
  {\bibinfo  {journal} {Nature communications}\ }\textbf {\bibinfo {volume}
  {4}},\ \bibinfo {pages} {1} (\bibinfo {year} {2013})}\BibitemShut {NoStop}%
\bibitem [{\citenamefont {Xiong}\ \emph {et~al.}(2017)\citenamefont {Xiong},
  \citenamefont {Gan},\ and\ \citenamefont {Wu}}]{xiong2017kuznetsov}%
  \BibitemOpen
  \bibfield  {author} {\bibinfo {author} {\bibfnamefont {H.}~\bibnamefont
  {Xiong}}, \bibinfo {author} {\bibfnamefont {J.}~\bibnamefont {Gan}},\ and\
  \bibinfo {author} {\bibfnamefont {Y.}~\bibnamefont {Wu}},\ }\bibfield
  {title} {\bibinfo {title} {Kuznetsov-ma soliton dynamics based on the
  mechanical effect of light},\ }\href@noop {} {\bibfield  {journal} {\bibinfo
  {journal} {Physical review letters}\ }\textbf {\bibinfo {volume} {119}},\
  \bibinfo {pages} {153901} (\bibinfo {year} {2017})}\BibitemShut {NoStop}%
\bibitem [{\citenamefont {Bao}\ \emph {et~al.}(2019)\citenamefont {Bao},
  \citenamefont {Cooper}, \citenamefont {Rowley}, \citenamefont {Di~Lauro},
  \citenamefont {Gongora}, \citenamefont {Chu}, \citenamefont {Little},
  \citenamefont {Oppo}, \citenamefont {Morandotti}, \citenamefont {Moss} \emph
  {et~al.}}]{bao2019laser}%
  \BibitemOpen
  \bibfield  {author} {\bibinfo {author} {\bibfnamefont {H.}~\bibnamefont
  {Bao}}, \bibinfo {author} {\bibfnamefont {A.}~\bibnamefont {Cooper}},
  \bibinfo {author} {\bibfnamefont {M.}~\bibnamefont {Rowley}}, \bibinfo
  {author} {\bibfnamefont {L.}~\bibnamefont {Di~Lauro}}, \bibinfo {author}
  {\bibfnamefont {J.~S.~T.}\ \bibnamefont {Gongora}}, \bibinfo {author}
  {\bibfnamefont {S.~T.}\ \bibnamefont {Chu}}, \bibinfo {author} {\bibfnamefont
  {B.~E.}\ \bibnamefont {Little}}, \bibinfo {author} {\bibfnamefont {G.-L.}\
  \bibnamefont {Oppo}}, \bibinfo {author} {\bibfnamefont {R.}~\bibnamefont
  {Morandotti}}, \bibinfo {author} {\bibfnamefont {D.~J.}\ \bibnamefont
  {Moss}}, \emph {et~al.},\ }\bibfield  {title} {\bibinfo {title} {Laser
  cavity-soliton microcombs},\ }\href@noop {} {\bibfield  {journal} {\bibinfo
  {journal} {Nature Photonics}\ }\textbf {\bibinfo {volume} {13}},\ \bibinfo
  {pages} {384} (\bibinfo {year} {2019})}\BibitemShut {NoStop}%
\bibitem [{\citenamefont {Xia}\ \emph {et~al.}(2021)\citenamefont {Xia},
  \citenamefont {Kaltsas}, \citenamefont {Song}, \citenamefont {Komis},
  \citenamefont {Xu}, \citenamefont {Szameit}, \citenamefont {Buljan},
  \citenamefont {Makris},\ and\ \citenamefont {Chen}}]{xia2021nonlinear}%
  \BibitemOpen
  \bibfield  {author} {\bibinfo {author} {\bibfnamefont {S.}~\bibnamefont
  {Xia}}, \bibinfo {author} {\bibfnamefont {D.}~\bibnamefont {Kaltsas}},
  \bibinfo {author} {\bibfnamefont {D.}~\bibnamefont {Song}}, \bibinfo {author}
  {\bibfnamefont {I.}~\bibnamefont {Komis}}, \bibinfo {author} {\bibfnamefont
  {J.}~\bibnamefont {Xu}}, \bibinfo {author} {\bibfnamefont {A.}~\bibnamefont
  {Szameit}}, \bibinfo {author} {\bibfnamefont {H.}~\bibnamefont {Buljan}},
  \bibinfo {author} {\bibfnamefont {K.~G.}\ \bibnamefont {Makris}},\ and\
  \bibinfo {author} {\bibfnamefont {Z.}~\bibnamefont {Chen}},\ }\bibfield
  {title} {\bibinfo {title} {Nonlinear tuning of pt symmetry and non-hermitian
  topological states},\ }\href@noop {} {\bibfield  {journal} {\bibinfo
  {journal} {Science}\ }\textbf {\bibinfo {volume} {372}},\ \bibinfo {pages}
  {72} (\bibinfo {year} {2021})}\BibitemShut {NoStop}%
\bibitem [{\citenamefont {Zhang}\ \emph {et~al.}(2021)\citenamefont {Zhang},
  \citenamefont {Fu}, \citenamefont {Zhu}, \citenamefont {Fan}, \citenamefont
  {Chen}, \citenamefont {Wang}, \citenamefont {Liu}, \citenamefont {Baltuska},
  \citenamefont {Jin}, \citenamefont {Tian} \emph
  {et~al.}}]{zhang2021solitary}%
  \BibitemOpen
  \bibfield  {author} {\bibinfo {author} {\bibfnamefont {S.}~\bibnamefont
  {Zhang}}, \bibinfo {author} {\bibfnamefont {Z.}~\bibnamefont {Fu}}, \bibinfo
  {author} {\bibfnamefont {B.}~\bibnamefont {Zhu}}, \bibinfo {author}
  {\bibfnamefont {G.}~\bibnamefont {Fan}}, \bibinfo {author} {\bibfnamefont
  {Y.}~\bibnamefont {Chen}}, \bibinfo {author} {\bibfnamefont {S.}~\bibnamefont
  {Wang}}, \bibinfo {author} {\bibfnamefont {Y.}~\bibnamefont {Liu}}, \bibinfo
  {author} {\bibfnamefont {A.}~\bibnamefont {Baltuska}}, \bibinfo {author}
  {\bibfnamefont {C.}~\bibnamefont {Jin}}, \bibinfo {author} {\bibfnamefont
  {C.}~\bibnamefont {Tian}}, \emph {et~al.},\ }\bibfield  {title} {\bibinfo
  {title} {Solitary beam propagation in periodic layered kerr media enables
  high-efficiency pulse compression and mode self-cleaning},\ }\href@noop {}
  {\bibfield  {journal} {\bibinfo  {journal} {Light: Science \& Applications}\
  }\textbf {\bibinfo {volume} {10}},\ \bibinfo {pages} {1} (\bibinfo {year}
  {2021})}\BibitemShut {NoStop}%
\bibitem [{\citenamefont {Krupa}\ \emph {et~al.}(2016)\citenamefont {Krupa},
  \citenamefont {Tonello}, \citenamefont {Barth{\'e}l{\'e}my}, \citenamefont
  {Couderc}, \citenamefont {Shalaby}, \citenamefont {Bendahmane}, \citenamefont
  {Millot},\ and\ \citenamefont {Wabnitz}}]{krupa2016observation}%
  \BibitemOpen
  \bibfield  {author} {\bibinfo {author} {\bibfnamefont {K.}~\bibnamefont
  {Krupa}}, \bibinfo {author} {\bibfnamefont {A.}~\bibnamefont {Tonello}},
  \bibinfo {author} {\bibfnamefont {A.}~\bibnamefont {Barth{\'e}l{\'e}my}},
  \bibinfo {author} {\bibfnamefont {V.}~\bibnamefont {Couderc}}, \bibinfo
  {author} {\bibfnamefont {B.~M.}\ \bibnamefont {Shalaby}}, \bibinfo {author}
  {\bibfnamefont {A.}~\bibnamefont {Bendahmane}}, \bibinfo {author}
  {\bibfnamefont {G.}~\bibnamefont {Millot}},\ and\ \bibinfo {author}
  {\bibfnamefont {S.}~\bibnamefont {Wabnitz}},\ }\bibfield  {title} {\bibinfo
  {title} {Observation of geometric parametric instability induced by the
  periodic spatial self-imaging of multimode waves},\ }\href@noop {} {\bibfield
   {journal} {\bibinfo  {journal} {Physical review letters}\ }\textbf {\bibinfo
  {volume} {116}},\ \bibinfo {pages} {183901} (\bibinfo {year}
  {2016})}\BibitemShut {NoStop}%
\bibitem [{\citenamefont {Longhi}(2003)}]{longhi2003modulational}%
  \BibitemOpen
  \bibfield  {author} {\bibinfo {author} {\bibfnamefont {S.}~\bibnamefont
  {Longhi}},\ }\bibfield  {title} {\bibinfo {title} {Modulational instability
  and space time dynamics in nonlinear parabolic-index optical fibers},\
  }\href@noop {} {\bibfield  {journal} {\bibinfo  {journal} {Optics letters}\
  }\textbf {\bibinfo {volume} {28}},\ \bibinfo {pages} {2363} (\bibinfo {year}
  {2003})}\BibitemShut {NoStop}%
\bibitem [{\citenamefont {Krupa}\ \emph {et~al.}(2017)\citenamefont {Krupa},
  \citenamefont {Tonello}, \citenamefont {Shalaby}, \citenamefont {Fabert},
  \citenamefont {Barth{\'e}l{\'e}my}, \citenamefont {Millot}, \citenamefont
  {Wabnitz},\ and\ \citenamefont {Couderc}}]{krupa2017spatial}%
  \BibitemOpen
  \bibfield  {author} {\bibinfo {author} {\bibfnamefont {K.}~\bibnamefont
  {Krupa}}, \bibinfo {author} {\bibfnamefont {A.}~\bibnamefont {Tonello}},
  \bibinfo {author} {\bibfnamefont {B.~M.}\ \bibnamefont {Shalaby}}, \bibinfo
  {author} {\bibfnamefont {M.}~\bibnamefont {Fabert}}, \bibinfo {author}
  {\bibfnamefont {A.}~\bibnamefont {Barth{\'e}l{\'e}my}}, \bibinfo {author}
  {\bibfnamefont {G.}~\bibnamefont {Millot}}, \bibinfo {author} {\bibfnamefont
  {S.}~\bibnamefont {Wabnitz}},\ and\ \bibinfo {author} {\bibfnamefont
  {V.}~\bibnamefont {Couderc}},\ }\bibfield  {title} {\bibinfo {title} {Spatial
  beam self-cleaning in multimode fibres},\ }\href@noop {} {\bibfield
  {journal} {\bibinfo  {journal} {Nature Photonics}\ }\textbf {\bibinfo
  {volume} {11}},\ \bibinfo {pages} {237} (\bibinfo {year} {2017})}\BibitemShut
  {NoStop}%
\bibitem [{\citenamefont {Wright}\ \emph {et~al.}(2017)\citenamefont {Wright},
  \citenamefont {Christodoulides},\ and\ \citenamefont
  {Wise}}]{wright2017spatiotemporal}%
  \BibitemOpen
  \bibfield  {author} {\bibinfo {author} {\bibfnamefont {L.~G.}\ \bibnamefont
  {Wright}}, \bibinfo {author} {\bibfnamefont {D.~N.}\ \bibnamefont
  {Christodoulides}},\ and\ \bibinfo {author} {\bibfnamefont {F.~W.}\
  \bibnamefont {Wise}},\ }\bibfield  {title} {\bibinfo {title} {Spatiotemporal
  mode-locking in multimode fiber lasers},\ }\href@noop {} {\bibfield
  {journal} {\bibinfo  {journal} {Science}\ }\textbf {\bibinfo {volume}
  {358}},\ \bibinfo {pages} {94} (\bibinfo {year} {2017})}\BibitemShut
  {NoStop}%
\bibitem [{\citenamefont {Lopez-Galmiche}\ \emph {et~al.}(2016)\citenamefont
  {Lopez-Galmiche}, \citenamefont {Eznaveh}, \citenamefont {Eftekhar},
  \citenamefont {Lopez}, \citenamefont {Wright}, \citenamefont {Wise},
  \citenamefont {Christodoulides},\ and\ \citenamefont
  {Correa}}]{lopez2016visible}%
  \BibitemOpen
  \bibfield  {author} {\bibinfo {author} {\bibfnamefont {G.}~\bibnamefont
  {Lopez-Galmiche}}, \bibinfo {author} {\bibfnamefont {Z.~S.}\ \bibnamefont
  {Eznaveh}}, \bibinfo {author} {\bibfnamefont {M.}~\bibnamefont {Eftekhar}},
  \bibinfo {author} {\bibfnamefont {J.~A.}\ \bibnamefont {Lopez}}, \bibinfo
  {author} {\bibfnamefont {L.}~\bibnamefont {Wright}}, \bibinfo {author}
  {\bibfnamefont {F.}~\bibnamefont {Wise}}, \bibinfo {author} {\bibfnamefont
  {D.}~\bibnamefont {Christodoulides}},\ and\ \bibinfo {author} {\bibfnamefont
  {R.~A.}\ \bibnamefont {Correa}},\ }\bibfield  {title} {\bibinfo {title}
  {Visible supercontinuum generation in a graded index multimode fiber pumped
  at 1064 nm},\ }\href@noop {} {\bibfield  {journal} {\bibinfo  {journal}
  {Optics letters}\ }\textbf {\bibinfo {volume} {41}},\ \bibinfo {pages} {2553}
  (\bibinfo {year} {2016})}\BibitemShut {NoStop}%
\bibitem [{\citenamefont {Wright}\ \emph {et~al.}(2015)\citenamefont {Wright},
  \citenamefont {Christodoulides},\ and\ \citenamefont
  {Wise}}]{wright2015controllable}%
  \BibitemOpen
  \bibfield  {author} {\bibinfo {author} {\bibfnamefont {L.~G.}\ \bibnamefont
  {Wright}}, \bibinfo {author} {\bibfnamefont {D.~N.}\ \bibnamefont
  {Christodoulides}},\ and\ \bibinfo {author} {\bibfnamefont {F.~W.}\
  \bibnamefont {Wise}},\ }\bibfield  {title} {\bibinfo {title} {Controllable
  spatiotemporal nonlinear effects in multimode fibres},\ }\href@noop {}
  {\bibfield  {journal} {\bibinfo  {journal} {Nature photonics}\ }\textbf
  {\bibinfo {volume} {9}},\ \bibinfo {pages} {306} (\bibinfo {year}
  {2015})}\BibitemShut {NoStop}%
\bibitem [{\citenamefont {Ramos}\ \emph {et~al.}(2020)\citenamefont {Ramos},
  \citenamefont {Fern{\'a}ndez-Alc{\'a}zar}, \citenamefont {Kottos},\ and\
  \citenamefont {Shapiro}}]{ramos2020optical}%
  \BibitemOpen
  \bibfield  {author} {\bibinfo {author} {\bibfnamefont {A.}~\bibnamefont
  {Ramos}}, \bibinfo {author} {\bibfnamefont {L.}~\bibnamefont
  {Fern{\'a}ndez-Alc{\'a}zar}}, \bibinfo {author} {\bibfnamefont
  {T.}~\bibnamefont {Kottos}},\ and\ \bibinfo {author} {\bibfnamefont
  {B.}~\bibnamefont {Shapiro}},\ }\bibfield  {title} {\bibinfo {title} {Optical
  phase transitions in photonic networks: a spin-system formulation},\
  }\href@noop {} {\bibfield  {journal} {\bibinfo  {journal} {Physical Review
  X}\ }\textbf {\bibinfo {volume} {10}},\ \bibinfo {pages} {031024} (\bibinfo
  {year} {2020})}\BibitemShut {NoStop}%
\bibitem [{\citenamefont {Shi}\ \emph {et~al.}(2021)\citenamefont {Shi},
  \citenamefont {Kottos},\ and\ \citenamefont {Shapiro}}]{shi2021controlling}%
  \BibitemOpen
  \bibfield  {author} {\bibinfo {author} {\bibfnamefont {C.}~\bibnamefont
  {Shi}}, \bibinfo {author} {\bibfnamefont {T.}~\bibnamefont {Kottos}},\ and\
  \bibinfo {author} {\bibfnamefont {B.}~\bibnamefont {Shapiro}},\ }\bibfield
  {title} {\bibinfo {title} {Controlling optical beam thermalization via
  band-gap engineering},\ }\href@noop {} {\bibfield  {journal} {\bibinfo
  {journal} {arXiv preprint arXiv:2108.05072}\ } (\bibinfo {year}
  {2021})}\BibitemShut {NoStop}%
\bibitem [{\citenamefont {Parto}\ \emph {et~al.}(2019)\citenamefont {Parto},
  \citenamefont {Wu}, \citenamefont {Jung}, \citenamefont {Makris},\ and\
  \citenamefont {Christodoulides}}]{parto2019thermodynamic}%
  \BibitemOpen
  \bibfield  {author} {\bibinfo {author} {\bibfnamefont {M.}~\bibnamefont
  {Parto}}, \bibinfo {author} {\bibfnamefont {F.~O.}\ \bibnamefont {Wu}},
  \bibinfo {author} {\bibfnamefont {P.~S.}\ \bibnamefont {Jung}}, \bibinfo
  {author} {\bibfnamefont {K.}~\bibnamefont {Makris}},\ and\ \bibinfo {author}
  {\bibfnamefont {D.~N.}\ \bibnamefont {Christodoulides}},\ }\bibfield  {title}
  {\bibinfo {title} {Thermodynamic conditions governing the optical temperature
  and chemical potential in nonlinear highly multimoded photonic systems},\
  }\href@noop {} {\bibfield  {journal} {\bibinfo  {journal} {Optics letters}\
  }\textbf {\bibinfo {volume} {44}},\ \bibinfo {pages} {3936} (\bibinfo {year}
  {2019})}\BibitemShut {NoStop}%
\bibitem [{\citenamefont {Wu}\ \emph {et~al.}(2019)\citenamefont {Wu},
  \citenamefont {Hassan},\ and\ \citenamefont
  {Christodoulides}}]{wu2019thermodynamic}%
  \BibitemOpen
  \bibfield  {author} {\bibinfo {author} {\bibfnamefont {F.~O.}\ \bibnamefont
  {Wu}}, \bibinfo {author} {\bibfnamefont {A.~U.}\ \bibnamefont {Hassan}},\
  and\ \bibinfo {author} {\bibfnamefont {D.~N.}\ \bibnamefont
  {Christodoulides}},\ }\bibfield  {title} {\bibinfo {title} {Thermodynamic
  theory of highly multimoded nonlinear optical systems},\ }\href@noop {}
  {\bibfield  {journal} {\bibinfo  {journal} {Nature Photonics}\ }\textbf
  {\bibinfo {volume} {13}},\ \bibinfo {pages} {776} (\bibinfo {year}
  {2019})}\BibitemShut {NoStop}%
\bibitem [{\citenamefont {Makris}\ \emph {et~al.}(2020)\citenamefont {Makris},
  \citenamefont {Wu}, \citenamefont {Jung},\ and\ \citenamefont
  {Christodoulides}}]{makris2020statistical}%
  \BibitemOpen
  \bibfield  {author} {\bibinfo {author} {\bibfnamefont {K.~G.}\ \bibnamefont
  {Makris}}, \bibinfo {author} {\bibfnamefont {F.~O.}\ \bibnamefont {Wu}},
  \bibinfo {author} {\bibfnamefont {P.~S.}\ \bibnamefont {Jung}},\ and\
  \bibinfo {author} {\bibfnamefont {D.~N.}\ \bibnamefont {Christodoulides}},\
  }\bibfield  {title} {\bibinfo {title} {Statistical mechanics of weakly
  nonlinear optical multimode gases},\ }\href@noop {} {\bibfield  {journal}
  {\bibinfo  {journal} {Optics letters}\ }\textbf {\bibinfo {volume} {45}},\
  \bibinfo {pages} {1651} (\bibinfo {year} {2020})}\BibitemShut {NoStop}%
\bibitem [{\citenamefont {Yablonovitch}(1987)}]{yablonovitch1987inhibited}%
  \BibitemOpen
  \bibfield  {author} {\bibinfo {author} {\bibfnamefont {E.}~\bibnamefont
  {Yablonovitch}},\ }\bibfield  {title} {\bibinfo {title} {Inhibited
  spontaneous emission in solid-state physics and electronics},\ }\href@noop {}
  {\bibfield  {journal} {\bibinfo  {journal} {Physical review letters}\
  }\textbf {\bibinfo {volume} {58}},\ \bibinfo {pages} {2059} (\bibinfo {year}
  {1987})}\BibitemShut {NoStop}%
\bibitem [{\citenamefont {Pertsch}\ \emph {et~al.}(1999)\citenamefont
  {Pertsch}, \citenamefont {Dannberg}, \citenamefont {Elflein}, \citenamefont
  {Br{\"a}uer},\ and\ \citenamefont {Lederer}}]{pertsch1999optical}%
  \BibitemOpen
  \bibfield  {author} {\bibinfo {author} {\bibfnamefont {T.}~\bibnamefont
  {Pertsch}}, \bibinfo {author} {\bibfnamefont {P.}~\bibnamefont {Dannberg}},
  \bibinfo {author} {\bibfnamefont {W.}~\bibnamefont {Elflein}}, \bibinfo
  {author} {\bibfnamefont {A.}~\bibnamefont {Br{\"a}uer}},\ and\ \bibinfo
  {author} {\bibfnamefont {F.}~\bibnamefont {Lederer}},\ }\bibfield  {title}
  {\bibinfo {title} {Optical bloch oscillations in temperature tuned waveguide
  arrays},\ }\href@noop {} {\bibfield  {journal} {\bibinfo  {journal} {Physical
  Review Letters}\ }\textbf {\bibinfo {volume} {83}},\ \bibinfo {pages} {4752}
  (\bibinfo {year} {1999})}\BibitemShut {NoStop}%
\bibitem [{\citenamefont {Peschel}\ \emph {et~al.}(1998)\citenamefont
  {Peschel}, \citenamefont {Pertsch},\ and\ \citenamefont
  {Lederer}}]{peschel1998optical}%
  \BibitemOpen
  \bibfield  {author} {\bibinfo {author} {\bibfnamefont {U.}~\bibnamefont
  {Peschel}}, \bibinfo {author} {\bibfnamefont {T.}~\bibnamefont {Pertsch}},\
  and\ \bibinfo {author} {\bibfnamefont {F.}~\bibnamefont {Lederer}},\
  }\bibfield  {title} {\bibinfo {title} {Optical bloch oscillations in
  waveguide arrays},\ }\href@noop {} {\bibfield  {journal} {\bibinfo  {journal}
  {Optics letters}\ }\textbf {\bibinfo {volume} {23}},\ \bibinfo {pages} {1701}
  (\bibinfo {year} {1998})}\BibitemShut {NoStop}%
\bibitem [{\citenamefont {Rechtsman}\ \emph {et~al.}(2013)\citenamefont
  {Rechtsman}, \citenamefont {Zeuner}, \citenamefont {Plotnik}, \citenamefont
  {Lumer}, \citenamefont {Podolsky}, \citenamefont {Dreisow}, \citenamefont
  {Nolte}, \citenamefont {Segev},\ and\ \citenamefont
  {Szameit}}]{rechtsman2013photonic}%
  \BibitemOpen
  \bibfield  {author} {\bibinfo {author} {\bibfnamefont {M.~C.}\ \bibnamefont
  {Rechtsman}}, \bibinfo {author} {\bibfnamefont {J.~M.}\ \bibnamefont
  {Zeuner}}, \bibinfo {author} {\bibfnamefont {Y.}~\bibnamefont {Plotnik}},
  \bibinfo {author} {\bibfnamefont {Y.}~\bibnamefont {Lumer}}, \bibinfo
  {author} {\bibfnamefont {D.}~\bibnamefont {Podolsky}}, \bibinfo {author}
  {\bibfnamefont {F.}~\bibnamefont {Dreisow}}, \bibinfo {author} {\bibfnamefont
  {S.}~\bibnamefont {Nolte}}, \bibinfo {author} {\bibfnamefont
  {M.}~\bibnamefont {Segev}},\ and\ \bibinfo {author} {\bibfnamefont
  {A.}~\bibnamefont {Szameit}},\ }\bibfield  {title} {\bibinfo {title}
  {Photonic floquet topological insulators},\ }\href@noop {} {\bibfield
  {journal} {\bibinfo  {journal} {Nature}\ }\textbf {\bibinfo {volume} {496}},\
  \bibinfo {pages} {196} (\bibinfo {year} {2013})}\BibitemShut {NoStop}%
\bibitem [{\citenamefont {Xia}\ \emph {et~al.}(2018)\citenamefont {Xia},
  \citenamefont {Ramachandran}, \citenamefont {Xia}, \citenamefont {Li},
  \citenamefont {Liu}, \citenamefont {Tang}, \citenamefont {Hu}, \citenamefont
  {Song}, \citenamefont {Xu}, \citenamefont {Leykam}, \citenamefont {Flach},\
  and\ \citenamefont {Chen}}]{xia2018unconventional}%
  \BibitemOpen
  \bibfield  {author} {\bibinfo {author} {\bibfnamefont {S.}~\bibnamefont
  {Xia}}, \bibinfo {author} {\bibfnamefont {A.}~\bibnamefont {Ramachandran}},
  \bibinfo {author} {\bibfnamefont {S.}~\bibnamefont {Xia}}, \bibinfo {author}
  {\bibfnamefont {D.}~\bibnamefont {Li}}, \bibinfo {author} {\bibfnamefont
  {X.}~\bibnamefont {Liu}}, \bibinfo {author} {\bibfnamefont {L.}~\bibnamefont
  {Tang}}, \bibinfo {author} {\bibfnamefont {Y.}~\bibnamefont {Hu}}, \bibinfo
  {author} {\bibfnamefont {D.}~\bibnamefont {Song}}, \bibinfo {author}
  {\bibfnamefont {J.}~\bibnamefont {Xu}}, \bibinfo {author} {\bibfnamefont
  {D.}~\bibnamefont {Leykam}}, \bibinfo {author} {\bibfnamefont
  {S.}~\bibnamefont {Flach}},\ and\ \bibinfo {author} {\bibfnamefont
  {Z.}~\bibnamefont {Chen}},\ }\bibfield  {title} {\bibinfo {title}
  {Unconventional flatband line states in photonic lieb lattices},\ }\href@noop
  {} {\bibfield  {journal} {\bibinfo  {journal} {Physical review letters}\
  }\textbf {\bibinfo {volume} {121}},\ \bibinfo {pages} {263902} (\bibinfo
  {year} {2018})}\BibitemShut {NoStop}%
\bibitem [{\citenamefont {Trompeter}\ \emph {et~al.}(2006)\citenamefont
  {Trompeter}, \citenamefont {Pertsch}, \citenamefont {Lederer}, \citenamefont
  {Michaelis}, \citenamefont {Streppel}, \citenamefont {Br{\"a}uer},\ and\
  \citenamefont {Peschel}}]{trompeter2006visual}%
  \BibitemOpen
  \bibfield  {author} {\bibinfo {author} {\bibfnamefont {H.}~\bibnamefont
  {Trompeter}}, \bibinfo {author} {\bibfnamefont {T.}~\bibnamefont {Pertsch}},
  \bibinfo {author} {\bibfnamefont {F.}~\bibnamefont {Lederer}}, \bibinfo
  {author} {\bibfnamefont {D.}~\bibnamefont {Michaelis}}, \bibinfo {author}
  {\bibfnamefont {U.}~\bibnamefont {Streppel}}, \bibinfo {author}
  {\bibfnamefont {A.}~\bibnamefont {Br{\"a}uer}},\ and\ \bibinfo {author}
  {\bibfnamefont {U.}~\bibnamefont {Peschel}},\ }\bibfield  {title} {\bibinfo
  {title} {Visual observation of zener tunneling},\ }\href@noop {} {\bibfield
  {journal} {\bibinfo  {journal} {Physical review letters}\ }\textbf {\bibinfo
  {volume} {96}},\ \bibinfo {pages} {023901} (\bibinfo {year}
  {2006})}\BibitemShut {NoStop}%
\bibitem [{\citenamefont {Szameit}\ \emph {et~al.}(2009)\citenamefont
  {Szameit}, \citenamefont {Garanovich}, \citenamefont {Heinrich},
  \citenamefont {Sukhorukov}, \citenamefont {Dreisow}, \citenamefont {Pertsch},
  \citenamefont {Nolte}, \citenamefont {T{\"u}nnermann},\ and\ \citenamefont
  {Kivshar}}]{szameit2009polychromatic}%
  \BibitemOpen
  \bibfield  {author} {\bibinfo {author} {\bibfnamefont {A.}~\bibnamefont
  {Szameit}}, \bibinfo {author} {\bibfnamefont {I.~L.}\ \bibnamefont
  {Garanovich}}, \bibinfo {author} {\bibfnamefont {M.}~\bibnamefont
  {Heinrich}}, \bibinfo {author} {\bibfnamefont {A.~A.}\ \bibnamefont
  {Sukhorukov}}, \bibinfo {author} {\bibfnamefont {F.}~\bibnamefont {Dreisow}},
  \bibinfo {author} {\bibfnamefont {T.}~\bibnamefont {Pertsch}}, \bibinfo
  {author} {\bibfnamefont {S.}~\bibnamefont {Nolte}}, \bibinfo {author}
  {\bibfnamefont {A.}~\bibnamefont {T{\"u}nnermann}},\ and\ \bibinfo {author}
  {\bibfnamefont {Y.~S.}\ \bibnamefont {Kivshar}},\ }\bibfield  {title}
  {\bibinfo {title} {Polychromatic dynamic localization in curved photonic
  lattices},\ }\href@noop {} {\bibfield  {journal} {\bibinfo  {journal} {Nature
  Physics}\ }\textbf {\bibinfo {volume} {5}},\ \bibinfo {pages} {271} (\bibinfo
  {year} {2009})}\BibitemShut {NoStop}%
\bibitem [{\citenamefont {Lederer}\ \emph {et~al.}(2008)\citenamefont
  {Lederer}, \citenamefont {Stegeman}, \citenamefont {Christodoulides},
  \citenamefont {Assanto}, \citenamefont {Segev},\ and\ \citenamefont
  {Silberberg}}]{lederer2008discrete}%
  \BibitemOpen
  \bibfield  {author} {\bibinfo {author} {\bibfnamefont {F.}~\bibnamefont
  {Lederer}}, \bibinfo {author} {\bibfnamefont {G.~I.}\ \bibnamefont
  {Stegeman}}, \bibinfo {author} {\bibfnamefont {D.~N.}\ \bibnamefont
  {Christodoulides}}, \bibinfo {author} {\bibfnamefont {G.}~\bibnamefont
  {Assanto}}, \bibinfo {author} {\bibfnamefont {M.}~\bibnamefont {Segev}},\
  and\ \bibinfo {author} {\bibfnamefont {Y.}~\bibnamefont {Silberberg}},\
  }\bibfield  {title} {\bibinfo {title} {Discrete solitons in optics},\
  }\href@noop {} {\bibfield  {journal} {\bibinfo  {journal} {Physics Reports}\
  }\textbf {\bibinfo {volume} {463}},\ \bibinfo {pages} {1} (\bibinfo {year}
  {2008})}\BibitemShut {NoStop}%
\bibitem [{\citenamefont {Onorato}\ \emph {et~al.}(2015)\citenamefont
  {Onorato}, \citenamefont {Vozella}, \citenamefont {Proment},\ and\
  \citenamefont {Lvov}}]{onorato2015route}%
  \BibitemOpen
  \bibfield  {author} {\bibinfo {author} {\bibfnamefont {M.}~\bibnamefont
  {Onorato}}, \bibinfo {author} {\bibfnamefont {L.}~\bibnamefont {Vozella}},
  \bibinfo {author} {\bibfnamefont {D.}~\bibnamefont {Proment}},\ and\ \bibinfo
  {author} {\bibfnamefont {Y.~V.}\ \bibnamefont {Lvov}},\ }\bibfield  {title}
  {\bibinfo {title} {Route to thermalization in the $\alpha$-fermi--pasta--ulam
  system},\ }\href@noop {} {\bibfield  {journal} {\bibinfo  {journal}
  {Proceedings of the National Academy of Sciences}\ }\textbf {\bibinfo
  {volume} {112}},\ \bibinfo {pages} {4208} (\bibinfo {year}
  {2015})}\BibitemShut {NoStop}%
\bibitem [{\citenamefont {Dostart}\ \emph {et~al.}(2020)\citenamefont
  {Dostart}, \citenamefont {Zhang}, \citenamefont {Khilo}, \citenamefont
  {Brand}, \citenamefont {Al~Qubaisi}, \citenamefont {Onural}, \citenamefont
  {Feldkhun}, \citenamefont {Wagner},\ and\ \citenamefont
  {Popovi{\'c}}}]{dostart2020serpentine}%
  \BibitemOpen
  \bibfield  {author} {\bibinfo {author} {\bibfnamefont {N.}~\bibnamefont
  {Dostart}}, \bibinfo {author} {\bibfnamefont {B.}~\bibnamefont {Zhang}},
  \bibinfo {author} {\bibfnamefont {A.}~\bibnamefont {Khilo}}, \bibinfo
  {author} {\bibfnamefont {M.}~\bibnamefont {Brand}}, \bibinfo {author}
  {\bibfnamefont {K.}~\bibnamefont {Al~Qubaisi}}, \bibinfo {author}
  {\bibfnamefont {D.}~\bibnamefont {Onural}}, \bibinfo {author} {\bibfnamefont
  {D.}~\bibnamefont {Feldkhun}}, \bibinfo {author} {\bibfnamefont {K.~H.}\
  \bibnamefont {Wagner}},\ and\ \bibinfo {author} {\bibfnamefont {M.~A.}\
  \bibnamefont {Popovi{\'c}}},\ }\bibfield  {title} {\bibinfo {title}
  {Serpentine optical phased arrays for scalable integrated photonic lidar beam
  steering},\ }\href@noop {} {\bibfield  {journal} {\bibinfo  {journal}
  {Optica}\ }\textbf {\bibinfo {volume} {7}},\ \bibinfo {pages} {726} (\bibinfo
  {year} {2020})}\BibitemShut {NoStop}%
\bibitem [{\citenamefont {Partovi}\ \emph {et~al.}(1999)\citenamefont
  {Partovi}, \citenamefont {Peale}, \citenamefont {Wuttig}, \citenamefont
  {Murray}, \citenamefont {Zydzik}, \citenamefont {Hopkins}, \citenamefont
  {Baldwin}, \citenamefont {Hobson}, \citenamefont {Wynn}, \citenamefont
  {Lopata} \emph {et~al.}}]{partovi1999high}%
  \BibitemOpen
  \bibfield  {author} {\bibinfo {author} {\bibfnamefont {A.}~\bibnamefont
  {Partovi}}, \bibinfo {author} {\bibfnamefont {D.}~\bibnamefont {Peale}},
  \bibinfo {author} {\bibfnamefont {M.}~\bibnamefont {Wuttig}}, \bibinfo
  {author} {\bibfnamefont {C.~A.}\ \bibnamefont {Murray}}, \bibinfo {author}
  {\bibfnamefont {G.}~\bibnamefont {Zydzik}}, \bibinfo {author} {\bibfnamefont
  {L.}~\bibnamefont {Hopkins}}, \bibinfo {author} {\bibfnamefont
  {K.}~\bibnamefont {Baldwin}}, \bibinfo {author} {\bibfnamefont {W.~S.}\
  \bibnamefont {Hobson}}, \bibinfo {author} {\bibfnamefont {J.}~\bibnamefont
  {Wynn}}, \bibinfo {author} {\bibfnamefont {J.}~\bibnamefont {Lopata}}, \emph
  {et~al.},\ }\bibfield  {title} {\bibinfo {title} {High-power laser light
  source for near-field optics and its application to high-density optical data
  storage},\ }\href@noop {} {\bibfield  {journal} {\bibinfo  {journal} {Applied
  Physics Letters}\ }\textbf {\bibinfo {volume} {75}},\ \bibinfo {pages} {1515}
  (\bibinfo {year} {1999})}\BibitemShut {NoStop}%
\bibitem [{\citenamefont {Wu}\ \emph {et~al.}(2020)\citenamefont {Wu},
  \citenamefont {Jung}, \citenamefont {Parto}, \citenamefont {Khajavikhan},\
  and\ \citenamefont {Christodoulides}}]{wu2020entropic}%
  \BibitemOpen
  \bibfield  {author} {\bibinfo {author} {\bibfnamefont {F.~O.}\ \bibnamefont
  {Wu}}, \bibinfo {author} {\bibfnamefont {P.~S.}\ \bibnamefont {Jung}},
  \bibinfo {author} {\bibfnamefont {M.}~\bibnamefont {Parto}}, \bibinfo
  {author} {\bibfnamefont {M.}~\bibnamefont {Khajavikhan}},\ and\ \bibinfo
  {author} {\bibfnamefont {D.~N.}\ \bibnamefont {Christodoulides}},\ }\bibfield
   {title} {\bibinfo {title} {Entropic thermodynamics of nonlinear photonic
  chain networks},\ }\href@noop {} {\bibfield  {journal} {\bibinfo  {journal}
  {Communications Physics}\ }\textbf {\bibinfo {volume} {3}},\ \bibinfo {pages}
  {1} (\bibinfo {year} {2020})}\BibitemShut {NoStop}%
\bibitem [{\citenamefont {Rumpf}(2007)}]{rumpf2007growth}%
  \BibitemOpen
  \bibfield  {author} {\bibinfo {author} {\bibfnamefont {B.}~\bibnamefont
  {Rumpf}},\ }\bibfield  {title} {\bibinfo {title} {Growth and erosion of a
  discrete breather interacting with rayleigh-jeans distributed phonons},\
  }\href@noop {} {\bibfield  {journal} {\bibinfo  {journal} {EPL (Europhysics
  Letters)}\ }\textbf {\bibinfo {volume} {78}},\ \bibinfo {pages} {26001}
  (\bibinfo {year} {2007})}\BibitemShut {NoStop}%
\bibitem [{\citenamefont {Klaers}\ \emph {et~al.}(2010)\citenamefont {Klaers},
  \citenamefont {Schmitt}, \citenamefont {Vewinger},\ and\ \citenamefont
  {Weitz}}]{klaers2010bose}%
  \BibitemOpen
  \bibfield  {author} {\bibinfo {author} {\bibfnamefont {J.}~\bibnamefont
  {Klaers}}, \bibinfo {author} {\bibfnamefont {J.}~\bibnamefont {Schmitt}},
  \bibinfo {author} {\bibfnamefont {F.}~\bibnamefont {Vewinger}},\ and\
  \bibinfo {author} {\bibfnamefont {M.}~\bibnamefont {Weitz}},\ }\bibfield
  {title} {\bibinfo {title} {Bose--einstein condensation of photons in an
  optical microcavity},\ }\href@noop {} {\bibfield  {journal} {\bibinfo
  {journal} {Nature}\ }\textbf {\bibinfo {volume} {468}},\ \bibinfo {pages}
  {545} (\bibinfo {year} {2010})}\BibitemShut {NoStop}%
\bibitem [{\citenamefont {Deng}\ \emph {et~al.}(2002)\citenamefont {Deng},
  \citenamefont {Weihs}, \citenamefont {Santori}, \citenamefont {Bloch},\ and\
  \citenamefont {Yamamoto}}]{deng2002condensation}%
  \BibitemOpen
  \bibfield  {author} {\bibinfo {author} {\bibfnamefont {H.}~\bibnamefont
  {Deng}}, \bibinfo {author} {\bibfnamefont {G.}~\bibnamefont {Weihs}},
  \bibinfo {author} {\bibfnamefont {C.}~\bibnamefont {Santori}}, \bibinfo
  {author} {\bibfnamefont {J.}~\bibnamefont {Bloch}},\ and\ \bibinfo {author}
  {\bibfnamefont {Y.}~\bibnamefont {Yamamoto}},\ }\bibfield  {title} {\bibinfo
  {title} {Condensation of semiconductor microcavity exciton polaritons},\
  }\href@noop {} {\bibfield  {journal} {\bibinfo  {journal} {Science}\ }\textbf
  {\bibinfo {volume} {298}},\ \bibinfo {pages} {199} (\bibinfo {year}
  {2002})}\BibitemShut {NoStop}%
\bibitem [{\citenamefont {Kasprzak}\ \emph {et~al.}(2006)\citenamefont
  {Kasprzak}, \citenamefont {Richard}, \citenamefont {Kundermann},
  \citenamefont {Baas}, \citenamefont {Jeambrun}, \citenamefont {Keeling},
  \citenamefont {Marchetti}, \citenamefont {Szyma{\'n}ska}, \citenamefont
  {Andr{\'e}}, \citenamefont {Staehli} \emph {et~al.}}]{kasprzak2006bose}%
  \BibitemOpen
  \bibfield  {author} {\bibinfo {author} {\bibfnamefont {J.}~\bibnamefont
  {Kasprzak}}, \bibinfo {author} {\bibfnamefont {M.}~\bibnamefont {Richard}},
  \bibinfo {author} {\bibfnamefont {S.}~\bibnamefont {Kundermann}}, \bibinfo
  {author} {\bibfnamefont {A.}~\bibnamefont {Baas}}, \bibinfo {author}
  {\bibfnamefont {P.}~\bibnamefont {Jeambrun}}, \bibinfo {author}
  {\bibfnamefont {J.~M.~J.}\ \bibnamefont {Keeling}}, \bibinfo {author}
  {\bibfnamefont {F.}~\bibnamefont {Marchetti}}, \bibinfo {author}
  {\bibfnamefont {M.}~\bibnamefont {Szyma{\'n}ska}}, \bibinfo {author}
  {\bibfnamefont {R.}~\bibnamefont {Andr{\'e}}}, \bibinfo {author}
  {\bibfnamefont {J.}~\bibnamefont {Staehli}}, \emph {et~al.},\ }\bibfield
  {title} {\bibinfo {title} {Bose--einstein condensation of exciton
  polaritons},\ }\href@noop {} {\bibfield  {journal} {\bibinfo  {journal}
  {Nature}\ }\textbf {\bibinfo {volume} {443}},\ \bibinfo {pages} {409}
  (\bibinfo {year} {2006})}\BibitemShut {NoStop}%
\bibitem [{\citenamefont {Jamadi}\ \emph {et~al.}(2016)\citenamefont {Jamadi},
  \citenamefont {R{\'e}veret}, \citenamefont {Mallet}, \citenamefont {Disseix},
  \citenamefont {M{\'e}dard}, \citenamefont {Mihailovic}, \citenamefont
  {Solnyshkov}, \citenamefont {Malpuech}, \citenamefont {Leymarie},
  \citenamefont {Lafosse}, \citenamefont {Bouchoule}, \citenamefont {Li},
  \citenamefont {Leroux}, \citenamefont {Semond},\ and\ \citenamefont
  {Zuniga-Perez}}]{jamadi2016polariton}%
  \BibitemOpen
  \bibfield  {author} {\bibinfo {author} {\bibfnamefont {O.}~\bibnamefont
  {Jamadi}}, \bibinfo {author} {\bibfnamefont {F.}~\bibnamefont {R{\'e}veret}},
  \bibinfo {author} {\bibfnamefont {E.}~\bibnamefont {Mallet}}, \bibinfo
  {author} {\bibfnamefont {P.}~\bibnamefont {Disseix}}, \bibinfo {author}
  {\bibfnamefont {F.}~\bibnamefont {M{\'e}dard}}, \bibinfo {author}
  {\bibfnamefont {M.}~\bibnamefont {Mihailovic}}, \bibinfo {author}
  {\bibfnamefont {D.}~\bibnamefont {Solnyshkov}}, \bibinfo {author}
  {\bibfnamefont {G.}~\bibnamefont {Malpuech}}, \bibinfo {author}
  {\bibfnamefont {J.}~\bibnamefont {Leymarie}}, \bibinfo {author}
  {\bibfnamefont {X.}~\bibnamefont {Lafosse}}, \bibinfo {author} {\bibfnamefont
  {S.}~\bibnamefont {Bouchoule}}, \bibinfo {author} {\bibfnamefont
  {F.}~\bibnamefont {Li}}, \bibinfo {author} {\bibfnamefont {M.}~\bibnamefont
  {Leroux}}, \bibinfo {author} {\bibfnamefont {F.}~\bibnamefont {Semond}},\
  and\ \bibinfo {author} {\bibfnamefont {J.}~\bibnamefont {Zuniga-Perez}},\
  }\bibfield  {title} {\bibinfo {title} {Polariton condensation phase diagram
  in wide-band-gap planar microcavities: Gan versus zno},\ }\href@noop {}
  {\bibfield  {journal} {\bibinfo  {journal} {Physical Review B}\ }\textbf
  {\bibinfo {volume} {93}},\ \bibinfo {pages} {115205} (\bibinfo {year}
  {2016})}\BibitemShut {NoStop}%
\bibitem [{\citenamefont {Sun}\ \emph {et~al.}(2017)\citenamefont {Sun},
  \citenamefont {Wen}, \citenamefont {Yoon}, \citenamefont {Liu}, \citenamefont
  {Steger}, \citenamefont {Pfeiffer}, \citenamefont {West}, \citenamefont
  {Snoke},\ and\ \citenamefont {Nelson}}]{sun2017bose}%
  \BibitemOpen
  \bibfield  {author} {\bibinfo {author} {\bibfnamefont {Y.}~\bibnamefont
  {Sun}}, \bibinfo {author} {\bibfnamefont {P.}~\bibnamefont {Wen}}, \bibinfo
  {author} {\bibfnamefont {Y.}~\bibnamefont {Yoon}}, \bibinfo {author}
  {\bibfnamefont {G.}~\bibnamefont {Liu}}, \bibinfo {author} {\bibfnamefont
  {M.}~\bibnamefont {Steger}}, \bibinfo {author} {\bibfnamefont {L.~N.}\
  \bibnamefont {Pfeiffer}}, \bibinfo {author} {\bibfnamefont {K.}~\bibnamefont
  {West}}, \bibinfo {author} {\bibfnamefont {D.~W.}\ \bibnamefont {Snoke}},\
  and\ \bibinfo {author} {\bibfnamefont {K.~A.}\ \bibnamefont {Nelson}},\
  }\bibfield  {title} {\bibinfo {title} {Bose-einstein condensation of
  long-lifetime polaritons in thermal equilibrium},\ }\href@noop {} {\bibfield
  {journal} {\bibinfo  {journal} {Physical review letters}\ }\textbf {\bibinfo
  {volume} {118}},\ \bibinfo {pages} {016602} (\bibinfo {year}
  {2017})}\BibitemShut {NoStop}%
\bibitem [{\citenamefont {Walker}\ \emph {et~al.}(2018)\citenamefont {Walker},
  \citenamefont {Flatten}, \citenamefont {Hesten}, \citenamefont {Mintert},
  \citenamefont {Hunger}, \citenamefont {Trichet}, \citenamefont {Smith},\ and\
  \citenamefont {Nyman}}]{walker2018driven}%
  \BibitemOpen
  \bibfield  {author} {\bibinfo {author} {\bibfnamefont {B.~T.}\ \bibnamefont
  {Walker}}, \bibinfo {author} {\bibfnamefont {L.~C.}\ \bibnamefont {Flatten}},
  \bibinfo {author} {\bibfnamefont {H.~J.}\ \bibnamefont {Hesten}}, \bibinfo
  {author} {\bibfnamefont {F.}~\bibnamefont {Mintert}}, \bibinfo {author}
  {\bibfnamefont {D.}~\bibnamefont {Hunger}}, \bibinfo {author} {\bibfnamefont
  {A.~A.}\ \bibnamefont {Trichet}}, \bibinfo {author} {\bibfnamefont {J.~M.}\
  \bibnamefont {Smith}},\ and\ \bibinfo {author} {\bibfnamefont {R.~A.}\
  \bibnamefont {Nyman}},\ }\bibfield  {title} {\bibinfo {title}
  {Driven-dissipative non-equilibrium bose--einstein condensation of less than
  ten photons},\ }\href@noop {} {\bibfield  {journal} {\bibinfo  {journal}
  {Nature Physics}\ }\textbf {\bibinfo {volume} {14}},\ \bibinfo {pages} {1173}
  (\bibinfo {year} {2018})}\BibitemShut {NoStop}%
\bibitem [{\citenamefont {Boyd}\ and\ \citenamefont
  {Masters}(2008)}]{boyd2008nonlinear}%
  \BibitemOpen
  \bibfield  {author} {\bibinfo {author} {\bibfnamefont {R.}~\bibnamefont
  {Boyd}}\ and\ \bibinfo {author} {\bibfnamefont {B.}~\bibnamefont {Masters}},\
  }\href@noop {} {\emph {\bibinfo {title} {Nonlinear Optics 3rd edn (New York:
  Academic)}}}\ (\bibinfo  {publisher} {Elsevier},\ \bibinfo {year}
  {2008})\BibitemShut {NoStop}%
\bibitem [{\citenamefont {Zakharov}\ \emph {et~al.}(1985)\citenamefont
  {Zakharov}, \citenamefont {Musher},\ and\ \citenamefont
  {Rubenchik}}]{zakharov1985hamiltonian}%
  \BibitemOpen
  \bibfield  {author} {\bibinfo {author} {\bibfnamefont {V.~E.}\ \bibnamefont
  {Zakharov}}, \bibinfo {author} {\bibfnamefont {S.}~\bibnamefont {Musher}},\
  and\ \bibinfo {author} {\bibfnamefont {A.}~\bibnamefont {Rubenchik}},\
  }\bibfield  {title} {\bibinfo {title} {Hamiltonian approach to the
  description of non-linear plasma phenomena},\ }\href@noop {} {\bibfield
  {journal} {\bibinfo  {journal} {Physics reports}\ }\textbf {\bibinfo {volume}
  {129}},\ \bibinfo {pages} {285} (\bibinfo {year} {1985})}\BibitemShut
  {NoStop}%
\bibitem [{\citenamefont {Picozzi}\ \emph {et~al.}(2014)\citenamefont
  {Picozzi}, \citenamefont {Garnier}, \citenamefont {Hansson}, \citenamefont
  {Suret}, \citenamefont {Randoux}, \citenamefont {Millot},\ and\ \citenamefont
  {Christodoulides}}]{picozzi2014optical}%
  \BibitemOpen
  \bibfield  {author} {\bibinfo {author} {\bibfnamefont {A.}~\bibnamefont
  {Picozzi}}, \bibinfo {author} {\bibfnamefont {J.}~\bibnamefont {Garnier}},
  \bibinfo {author} {\bibfnamefont {T.}~\bibnamefont {Hansson}}, \bibinfo
  {author} {\bibfnamefont {P.}~\bibnamefont {Suret}}, \bibinfo {author}
  {\bibfnamefont {S.}~\bibnamefont {Randoux}}, \bibinfo {author} {\bibfnamefont
  {G.}~\bibnamefont {Millot}},\ and\ \bibinfo {author} {\bibfnamefont {D.~N.}\
  \bibnamefont {Christodoulides}},\ }\bibfield  {title} {\bibinfo {title}
  {Optical wave turbulence: Towards a unified nonequilibrium thermodynamic
  formulation of statistical nonlinear optics},\ }\href@noop {} {\bibfield
  {journal} {\bibinfo  {journal} {Physics Reports}\ }\textbf {\bibinfo {volume}
  {542}},\ \bibinfo {pages} {1} (\bibinfo {year} {2014})}\BibitemShut {NoStop}%
\bibitem [{\citenamefont {Butler}\ \emph {et~al.}(1984)\citenamefont {Butler},
  \citenamefont {Ackley},\ and\ \citenamefont {Botez}}]{butler1984coupled}%
  \BibitemOpen
  \bibfield  {author} {\bibinfo {author} {\bibfnamefont {J.}~\bibnamefont
  {Butler}}, \bibinfo {author} {\bibfnamefont {D.}~\bibnamefont {Ackley}},\
  and\ \bibinfo {author} {\bibfnamefont {D.}~\bibnamefont {Botez}},\ }\bibfield
   {title} {\bibinfo {title} {Coupled-mode analysis of phase-locked injection
  laser arrays},\ }\href@noop {} {\bibfield  {journal} {\bibinfo  {journal}
  {Applied Physics Letters}\ }\textbf {\bibinfo {volume} {44}},\ \bibinfo
  {pages} {293} (\bibinfo {year} {1984})}\BibitemShut {NoStop}%
\bibitem [{\citenamefont {Kapon}\ \emph {et~al.}(1984)\citenamefont {Kapon},
  \citenamefont {Katz},\ and\ \citenamefont {Yariv}}]{kapon1984supermode}%
  \BibitemOpen
  \bibfield  {author} {\bibinfo {author} {\bibfnamefont {E.}~\bibnamefont
  {Kapon}}, \bibinfo {author} {\bibfnamefont {J.}~\bibnamefont {Katz}},\ and\
  \bibinfo {author} {\bibfnamefont {A.}~\bibnamefont {Yariv}},\ }\bibfield
  {title} {\bibinfo {title} {Supermode analysis of phase-locked arrays of
  semiconductor lasers},\ }\href@noop {} {\bibfield  {journal} {\bibinfo
  {journal} {Optics letters}\ }\textbf {\bibinfo {volume} {9}},\ \bibinfo
  {pages} {125} (\bibinfo {year} {1984})}\BibitemShut {NoStop}%
\bibitem [{\citenamefont {Weisstein}(2002)}]{weisstein2002binomial}%
  \BibitemOpen
  \bibfield  {author} {\bibinfo {author} {\bibfnamefont {E.~W.}\ \bibnamefont
  {Weisstein}},\ }\bibfield  {title} {\bibinfo {title} {Binomial sums},\
  }\href@noop {} {\bibfield  {journal} {\bibinfo  {journal}
  {http://mathworld.wolfram.com/BinomialSums.html}\ } (\bibinfo {year}
  {2002})}\BibitemShut {NoStop}%
\bibitem [{\citenamefont {Prudnikov}\ \emph {et~al.}(1986)\citenamefont
  {Prudnikov}, \citenamefont {Brychkov},\ and\ \citenamefont
  {Marichev}}]{prudnikov1986integrals}%
  \BibitemOpen
  \bibfield  {author} {\bibinfo {author} {\bibfnamefont {A.~P.}\ \bibnamefont
  {Prudnikov}}, \bibinfo {author} {\bibfnamefont {Y.~A.}\ \bibnamefont
  {Brychkov}},\ and\ \bibinfo {author} {\bibfnamefont {O.~I.}\ \bibnamefont
  {Marichev}},\ }\href@noop {} {\emph {\bibinfo {title} {Integrals and series.
  Volume 1. Elementary functions}}}\ (\bibinfo  {publisher} {CRC Press; 1st
  edition (January 1, 1986)},\ \bibinfo {year} {1986})\BibitemShut {NoStop}%
\bibitem [{\citenamefont {Gradshteyn}\ and\ \citenamefont
  {Ryzhik}(2014)}]{gradshteyn2014table}%
  \BibitemOpen
  \bibfield  {author} {\bibinfo {author} {\bibfnamefont {I.~S.}\ \bibnamefont
  {Gradshteyn}}\ and\ \bibinfo {author} {\bibfnamefont {I.~M.}\ \bibnamefont
  {Ryzhik}},\ }\href@noop {} {\emph {\bibinfo {title} {Table of integrals,
  series, and products}}}\ (\bibinfo  {publisher} {Academic press},\ \bibinfo
  {year} {2014})\BibitemShut {NoStop}%
\bibitem [{\citenamefont {Berger}\ and\ \citenamefont
  {Fekete}(1986)}]{berger1986far}%
  \BibitemOpen
  \bibfield  {author} {\bibinfo {author} {\bibfnamefont {J.}~\bibnamefont
  {Berger}}\ and\ \bibinfo {author} {\bibfnamefont {D.}~\bibnamefont
  {Fekete}},\ }\bibfield  {title} {\bibinfo {title} {Far-field analytical model
  for laser arrays and fitting with experimental results},\ }\href@noop {}
  {\bibfield  {journal} {\bibinfo  {journal} {Applied physics letters}\
  }\textbf {\bibinfo {volume} {49}},\ \bibinfo {pages} {605} (\bibinfo {year}
  {1986})}\BibitemShut {NoStop}%
\bibitem [{\citenamefont {Beyer}(1991)}]{beyer1991crc}%
  \BibitemOpen
  \bibfield  {author} {\bibinfo {author} {\bibfnamefont {W.~H.}\ \bibnamefont
  {Beyer}},\ }\bibfield  {title} {\bibinfo {title} {Crc standard mathematical
  tables and formulae},\ }\href@noop {} {\bibfield  {journal} {\bibinfo
  {journal} {Boca Raton}\ } (\bibinfo {year} {1991})}\BibitemShut {NoStop}%
\end{thebibliography}%
\providecommand{\noopsort}[1]{}\providecommand{\singleletter}[1]{#1}%

\end{document}